\documentclass[twocolumn]{aastex631}
\shorttitle{METIS imaging simulations for FU Ori-type objects}
\shortauthors{Takami et al.}
\graphicspath{{./}{figures/}}

\begin{document}
\bibliographystyle{astron}

\title{ELT-METIS imaging simulations for disks and envelopes associated with FU Ori-type objects}

\author[0000-0002-0786-7307]{Michihiro Takami}
\affiliation{Institute of Astronomy and Astrophysics, Academia Sinica, 
11F of Astronomy-Mathematics Building,
No.1, Sec. 4, Roosevelt Rd., Taipei 106216, Taiwan, R.O.C.}

\author[0000-0002-6717-1977]{Gilles Otten}
\affiliation{Institute of Astronomy and Astrophysics, Academia Sinica, 
11F of Astronomy-Mathematics Building,
No.1, Sec. 4, Roosevelt Rd., Taipei 106216, Taiwan, R.O.C.}


\author[0000-0002-4006-6237]{Olivier Absil}
\affiliation{STAR Institute, Universit\'e de Li\`ege, All\'ee du Six Ao\^ut 19c, B-4000 Li\`ege, Belgium}

\author[0000-0003-0150-4430]{Christian Delacroix}
\affiliation{STAR Institute, Universit\'e de Li\`ege, All\'ee du Six Ao\^ut 19c, B-4000 Li\`ege, Belgium}

\author{Jennifer L. Karr}
\affiliation{Institute of Astronomy and Astrophysics, Academia Sinica, 
11F of Astronomy-Mathematics Building,
No.1, Sec. 4, Roosevelt Rd., Taipei 106216, Taiwan, R.O.C.}

\author[0000-0001-6491-1901]{Shiang-Yu Wang}
\affiliation{Institute of Astronomy and Astrophysics, Academia Sinica, 
11F of Astronomy-Mathematics Building,
No.1, Sec. 4, Roosevelt Rd., Taipei 106216, Taiwan, R.O.C.}



\begin{abstract}
We investigate the detectability of extended mid-infrared (MIR) emission associated with FU-Ori type objects (FUors) using the METIS coronagraphs on the 39-m Extremely Large Telescope (ELT).
The imaging simulations were made for three representative filters ($\lambda$=3.8, 4.8, and 11.3 \micron) of the METIS instrument.
We demonstrate that the detectability of the extended MIR emission using these coronagraphs is highly dependent on the uncertain nature of the central FUor and its circumstellar environment in various contexts{. These contexts are: (A)} whether the central radiation source is either a flat self-luminous accretion disk or a star at near-infrared (NIR) wavelengths{, (B)} the size of the accretion disk for the bright central MIR emission at milliarcsecond scales{, (C)} whether the extended emission is due to either an optically thick disk or an optically thin envelope{, and (D)} dust grain models.
Observations at $\lambda$=3.8 \micron~will allow us to detect the extended emission in many cases, while the number of cases with detection 
{may} significantly decrease
toward longer wavelengths due to the fainter nature of the extended emission and high thermal background noise. In some cases, the presence of a binary companion can significantly hamper detections of the extended MIR emission. NIR and MIR imaging observations at existing 8-m class telescopes, prior to the METIS observations, will be useful for (1) reducing the many model uncertainties and (2) searching for binary companions associated with FUors, therefore determining the best observing strategy using METIS.
\end{abstract}

\keywords{
Methods: observational
--
Techniques: image processing
--
Stars: protostars
---
Infrared: stars
}

\accepted{by PASP, February 28, 2025}


\section{Introduction} \label{sec:intro}

FU Orionis objects (hereafter FUors) are young stellar objects (YSOs) that undergo the most active and violent accretion outbursts. During each burst, the accretion rate rapidly increases by a factor of approximately 1000, and remains high for several decades or more.
It has been suggested that many low-mass YSOs experience FUor outbursts, but we miss most of them because of the small chance of capturing the events. Readers can refer to \citet{Hartmann96} and \citet{Audard14} for reviews of FUors.

Near-infrared (NIR; $\lambda$$\sim$2 \micron) imaging polarimetry
at high-angular resolutions ($\sim$0\farcs1) revealed complicated circumstellar structures associated with some FUors \citep{Liu16,Takami18,Laws20,Weber23,Zurlo24}. 
The observed extended emission at these wavelengths is due to scattering from circumstellar dust grains illuminated by the central source. \citet{Liu16} and \citet{Takami18} attributed the observed circumstellar structures to gravitationally unstable disks and trails of clump ejections in such disks.
This interpretation was corroborated by \citet{Weber23}, who used the Atacama Large Millimeter/submillimeter Array (ALMA) and revealed a fragmenting spiral structure in the disk associated with the FUor V960 Mon. 
Gravitational fragmentation induced by these instabilities may also induce the formation of planets and brown dwarfs at large orbital radii,  the presence of which conventional planet formation models cannot easily explain \citep[e.g.,][]{Boss03,Nayaksin10,Vorobyov13,Stamatellos15}.

However, the observed extended emission in the NIR may alternatively be explained by an extended envelope.
This explanation is corroborated by IR spectral energy distributions (SEDs) and millimeter emission, which indicate the presence of massive circumstellar envelopes toward some FUors \citep[e.g.,][]{Sandell01,Gramajo14}. Furthermore, \citet{Laws20} executed NIR imaging observations toward FU Ori, an archetypical FUor, and pointed out that the observed structures are similar to those of infalling gas toward some normal YSOs observed at millimeter wavelengths \citep[e.g.,][]{Yen14,Yen19}. If this is the case, the structures seen in the NIR images would provide valuable clues for understanding how the circumstellar disk is fed from the envelope, leading to accretion outbursts \citep[e.g.,][]{Hartmann96}.

Throughout, the circumstellar IR emission associated with FUors may hold keys to understanding protostellar evolution and planet formation, not only for FUors but also in a general context. Mid-IR (MIR; $\lambda$$\gtrsim$3 \micron) observations suffer less from circumstellar extinction than NIR wavelengths and therefore allow us to search for embedded disk emission if the extended NIR emission is due to a dusty envelope. Such studies have successfully been made for NIR imaging observations of the Herbig Ae star AB Aur, which revealed disk structures not observed in the optical images \citep{Fukagawa04,Hashimoto11}.
In turn, observations at longer wavelengths degrade the diffraction-limited angular resolution. This problem will be resolved by next-generation extremely large telescopes such as the Extremely Large Telescope (ELT, with a 39-m telescope diameter), the Giant Magellan Telescope (GMT, 25-m), and the Thirty Meter Telescope (TMT, 30-m).
{A 30-m telescope with adaptive optics (AO) will yield a diffraction-limited angular resolution of 25 mas at $\lambda$=3 $\micron$, the same as that for a 10-m telescope at $\lambda$=1 $\micron$.}

In particular, the Mid-infrared ELT Imager and Spectrograph (METIS) on ELT \citep{METIS22} will be a powerful tool for MIR imaging observations at 3-13 \micron. While the baseline design of METIS does not include an imaging polarimetry mode, its coronagraphs (high-contrast imaging elements, hereafter HCI) would be powerful tools for observing MIR circumstellar emission associated with bright central sources such as FUors. Furthermore, {AO} with the large telescope diameter of ELT will yield a diffraction-limited angular resolution of 20 mas at the shortest wavelengths, therefore improving the resolution of observations by up to a factor of $\sim$2 compared with the NIR imaging polarimetry to date {(mainly in the $H$-band at $\lambda$=1.65 $\micron$) } at 8-m class telescopes. The improved angular resolution may also be useful for better understanding the nature of the observed extended emission, for example, by revealing finer structures in gravitationally fragmenting disks \citep[e.g.,][]{Vorobyov15,Liu16,Dong16} or infalling envelopes \citep[e.g.,][]{Ginski21}.

In this paper we will present imaging simulations {of total intensity (Stokes $I$)} for FUor disks/envelopes of observations using HCI. 
We selected three representative METIS filters ($\lambda$=3.8, 4.8, and 11.3 \micron, respectively) for these simulations.
HCI will contain the Classical Vortex Coronagraph (CVC), the Ring Apodized Vortex Coronagraph (RAVC), and the Apodizing Phase Plate (APP) for observations at $\lambda$=3-5 \micron; and CVC for $\lambda = 8-13$ \micron. We used CVC and RAVC to observe the MIR emission extending over an arcsecond scale.
In Table \ref{tbl:inst}, we summarize the specifications for METIS and HCI. 
{The CVC yields a better throughput at the cost of a relatively modest starlight suppression, while the RAVC aims to provide the highest possible starlight suppression at the expense of throughput \citep[][see also Section \ref{sec:results:basic}]{Carlomagno20}.
}

\begin{table*}
\caption{Instrument parameters for METIS and HCI. \label{tbl:inst}}
\begin{tabular}{lccc}
\tableline\tableline
Parameter										& \multicolumn{3}{c}{Value}
\\ \tableline
Filter Name\tablenotemark{\scriptsize a}			& HCI-$L$ long			& CO ref					& GeoSnap N2 \\
Wavelength $\lambda$ (\micron) 				& 3.81				& 4.79					& 11.33 \\
Filter width $\Delta \lambda$ (\micron)			& 0.27				& 0.22					& 3.03 \\
{Detector} 								& \multicolumn{2}{c}{--- Hawaii2RG ---}	& {--- GeoSnap ---}  \\
~~~Pixel scale (mas)						& \multicolumn{2}{c}{ 5.47}						& 6.79 \\
~~~Minimum exposure (s)						& \multicolumn{2}{c}{ 0.04}						& 0.011 \\
~~~Saturation limit (e$^-$)					& \multicolumn{2}{c}{ 1$\times$10$^5$}				& $2.8 \times 10^6$		\\
~~~Read noise (e$^-$)						& \multicolumn{2}{c}{ 70}						& 300		\\
{Inner working angle (IWA, mas)}		& {25}				& {31}					& {73} \\
Flux for a zero-magnitude star (e$^-$ s$^{-1}$)\tablenotemark{\scriptsize \small b}		& 9.0$\times10^{10}$	& 2.5$\times10^{10}$	& 3.7$\times10^{10}$\\
Thermal background (e$^-$ s$^{-1}$ pixel$^{-1}$)\tablenotemark{\scriptsize b,c}	& 8.9$\times10^{4}$		& 6.7$\times10^{5}$		& 1.1$\times10^{8}$ \\
Off-axis Transmission, CVC\tablenotemark{\scriptsize d}		& 0.724				& 0.592				& 0.737\\
Off-axis Transmission, RAVC\tablenotemark{\scriptsize d}		& 0.334				& 0.274				& ---\tablenotemark{\scriptsize e}\\
Critical Exposure time, CVC\tablenotemark{\scriptsize e}		& 7.6$\times10^{-2}$		& (1.2$\times10^{-2}$)	& (1.1$\times10^{-3}$) \\
Critical Exposure time, RAVC\tablenotemark{\scriptsize e}		& 1.7$\times10^{-1}$		& 2.7$\times10^{-2}$		& ---\tablenotemark{\scriptsize \scriptsize f}\\
\tableline
\end{tabular} \\
\tablenotetext{a}{Labeled ``L", ``M", and ``N2" in HEEPS 1.0.0 for short to long wavelengths. The wavelengths for the former two approximately match the Johnson $L$ and $M$ filters, but the parameters for these three filters are optimized primarily for the observations of exoplanets.}
\tablenotetext{b}{Derived excluding the transmission of the coronagraph tabulated below.}
\tablenotetext{c}{See Section \ref{sec:method:heeps} for the assumed observing conditions.}
\tablenotetext{d}{For the coronagraph optics only.}
\tablenotetext{e}{Exposure time for which the thermal background photon noise is equal to the read noise. The values with brackets are significantly smaller than the detector minimum exposure time, implying that the noise is dominated by the photon noise for any possible exposure time for these cases.}
\tablenotetext{f}{No RAVC for the GeoSnap $N$2 band ($\lambda$=11.3 \micron).}
\end{table*}

In Section \ref{sec:method}, we summarize the method of our simulations, which consists of the following two parts: (1) calculations of the MIR intensity distributions for FUors 
(Section \ref{sec:method:extended}); and (2) use of the High-contrast ELT End-to-end Performance Simulator (HEEPS) to investigate the signal-to-noise of the extended MIR emission (Section \ref{sec:method:heeps}). 
We show our results in Section \ref{sec:results}. In Section \ref{sec:summary}, we summarize the results and discuss research strategies for the future.


\section{Method} \label{sec:method}

In Section \ref{sec:method:extended}, we describe our calculations for the extended MIR emission based on \citet{Takami23b} (Paper I).
In Section \ref{sec:method:cs}, we describe the models for the central compact disk responsible for the bright central emission in the MIR.
In Section \ref{sec:method:heeps}, we describe our simulations using HEEPS.


\subsection{Calculations for the extended MIR emission} \label{sec:method:extended}

In Paper I, we developed a semi-analytic method to calculate extended MIR emission for existing FUors with an order-of-magnitude accuracy. The calculations are made using
(1) the observed distribution of the polarized intensity (PI) in the $H$-band ($\lambda$=1.65 \micron);
(2) the observed SEDs at ultraviolet (UV) to MIR wavelengths; and
(3) assumed optical properties for dust grains in a disk or an envelope, which produces extended infrared emission.
Our method allows us to easily calculate the emission distributions for various cases{. These  cases are specifically:
(A)} when the central radiation source at UV to NIR wavelengths ($\lambda < 3$ \micron) is a flat compact self-luminous disk \citep[e.g.,][]{Hartmann96,Zhu08} or a star \citep[e.g.,][]{Herbig03,Elbakyan19}{,
(B)} when the infrared extended emission is associated with a disk or an envelope{, and
(C)} with different dust models for light scattering and thermal radiation from the extended disk or envelope.
{This semi-analytic method is complementary to full radiative transfer simulations, which offer more accurate calculations but only with specific dynamical models and significant computational time.}

We used some assumptions and simplification for the geometry of the extended disks/envelopes to derive the MIR images. 
This makes our calculations less accurate, but still with accuracies sufficient for our purpose, that is, determining whether the extended MIR emission is observable using HCI.
In Paper I, we performed comparisons with some numerical simulations, and these suggest that our semi-analytic method yields intensity distributions with an accuracy within a factor of 2.
However, this method does not include radiation heating in the inner disk edge or adiabatic heating, which potentially enhances thermal emission at $\lambda \sim 10$ $\micron$ (see also Section \ref{sec:summary} for future work).

\begin{figure}[ht!]
\centering
\includegraphics[width=9cm]{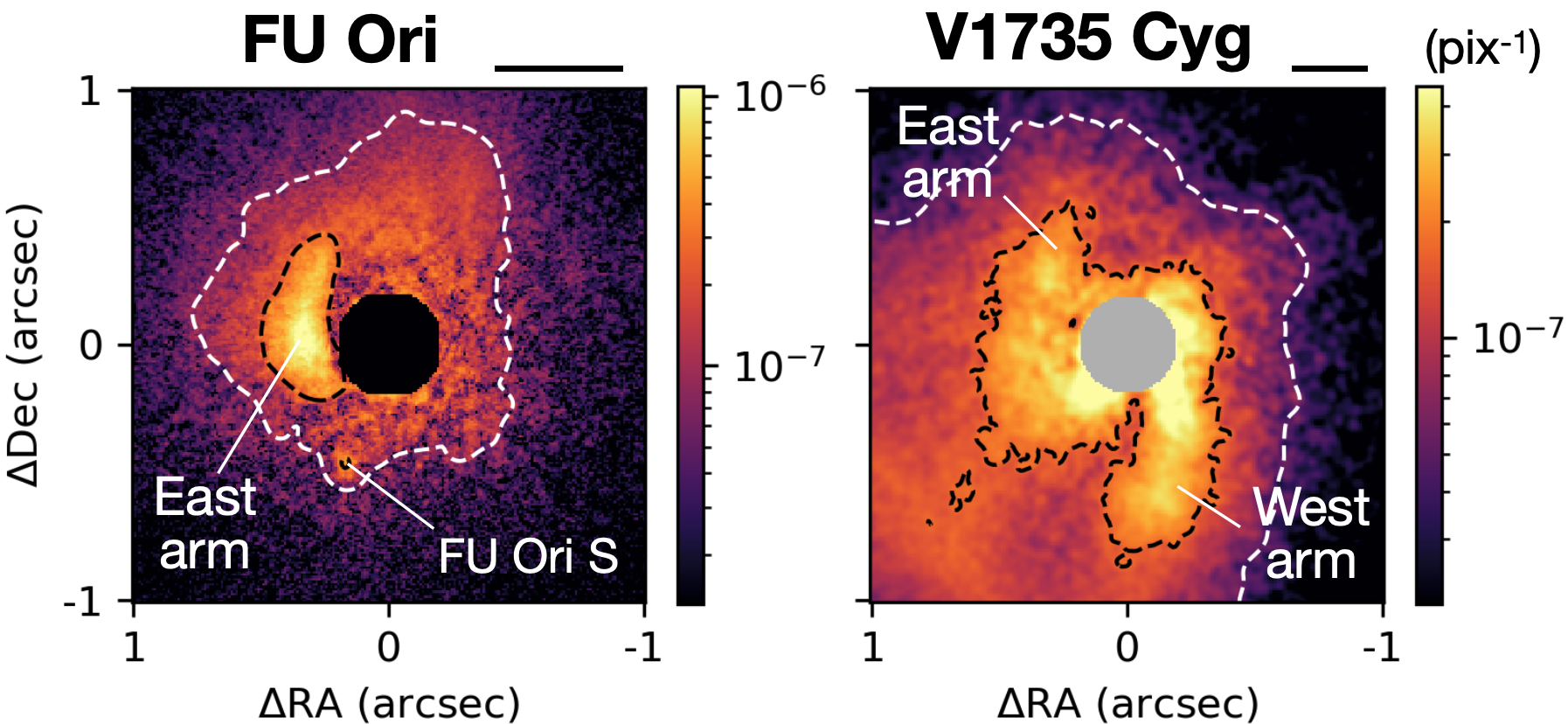}
\caption{
PI distribution in $H$ band ($\lambda$=1.65 \micron), $PI_{\mathrm{obs};H}$, for FU Ori and V1735 Cyg \citep{Takami18}. The intensity at each pixel (with a pixel scale of 9.5 mas for Subaru-HiCIAO) is normalized to the Stokes $I$ flux of the central source. North is up.
In each image, the central region is masked as we were not able to make a reliable measurement due to the central source being significantly brighter than the extended emission.
Labeled are arm-like features and the emission associated with the companion star FU Ori S.
The contours are shown with arbitrary scales (0.8$\times$10$^{-7}$ and 2.7$\times$10$^{-7}$  pixel$^{-1}$ for FU Ori; 0.5$\times$10$^{-7}$ and 1.7$\times$10$^{-7}$  pixel$^{-1}$ for V1735 Cyg) to indicate approximate emission distributions.
The black horizontal bar at the top-right of each panel indicates a spatial scale of 200 au (see Table \ref{tbl:targets} for the target distance).
\label{fig:PI_H}
}
\end{figure}

As in Paper I, we calculated the MIR emission for two FUors, FU Ori and V1735 Cyg.
We tabulate the key parameters for these objects in Table \ref{tbl:targets}.
Figure \ref{fig:PI_H} shows the images in $H$-band observed using the Subaru-HiCIAO instrument.
For each object, we set the intensity within 0\farcs2 of the central source to be zero as we were not able to make reliable measurement due to the central source being significantly brighter than the extended emission.
We note that this software mask is significantly larger than the inner working angle of the instrument tabulated in Table \ref{tbl:inst}.
In each panel, we also show a few contours to indicate approximate emission distributions. These contours will be used in Section \ref{sec:results:basic} to perform comparisons with the point-spread functions (PSFs) of the central source and their residuals for subtraction for the METIS observations.

For (2) described above, we used the SEDs described in Paper I.
In practice, V1735 Cyg cannot be observed from the ELT site due to its high declination. However, we will still use this target to investigate the detection of MIR emission associated with FUors in a more general context.

The infrared extended emission ($r > 10$ au), either observed or to be observed, must be due to dust scattering and thermal dust radiation in the disk or envelope \citep[Section 1; see also][]{Whitney03b,Whitney03a,Whitney04, Whitney13, Audard14}.
At UV to NIR wavelengths ($\lambda < 3$ \micron), dust grains in the extended disk or envelope are illuminated and heated by radiation from the central source, which is either a disk or a star as described above. At MIR wavelengths ($\lambda > 3$ \micron), the observed SEDs suggest that the radiation from the central source is dominated by the former \citep[][Paper I]{Liu16}. 

\begin{table*}
\caption{Targets\tablenotemark{\scriptsize a}. \label{tbl:targets}}
\begin{tabular}{lccccc}
\tableline\tableline
Target  		& Distance	& $L_\mathrm{*}$ 		& \multicolumn{3}{c}{$F_\mathrm{*}$ (W m$^{-2}$ \micron$^{-1}$)}	     \\
                		& (pc)            	& ($L_\sun$)			&  $\lambda$=3.8 \micron & 4.8 \micron & 11.3 \micron
\\ \tableline
FU Ori          	& 408$\pm$3   	& 1.0$\times$10$^2$     	& $1.1 \times 10^{-12}$ 	& $6.1 \times 10^{-13}$ 	& $1.1 \times 10^{-13}$  \\
V1735 Cyg       & 690$\pm$40 	& 43                                   & $3.2 \times 10^{-13}$ 	& $2.1 \times 10^{-13}$ 	& $3.3 \times 10^{-14}$ \\
\tableline
\end{tabular} \\
\tablenotetext{a}{Paper I}
\end{table*}

We approximated an extended disk with an optically thin atmosphere and an optically thin interior. For the disk, we then calculated emission via single scattering (the central source$\rightarrow$scattering in the disk atmosphere$\rightarrow$the observer), double scattering (the central source$\rightarrow$the first scattering in the disk atmosphere$\rightarrow$the second scattering in the disk interior$\rightarrow$the observer), and thermal emission from the disk atmosphere and the interior.
{
The light with more than two scatterings, which is not included in our calculations, would enhance the total intensity by only $\lesssim$5 \%.
This estimate is based on the fact that light even via double scattering contributes to the total intensity only up to $\sim$20 \% as described below.
}
For the extended envelope, we calculated the scattered and thermal emission assuming that the envelope is optically thin and therefore double scattering and self absorption are negligible. 

For the optical properties of dust grains, we used three models as for Paper I.
These dust models were originally developed for the interstellar medium without ice coating (`Dust1'), for a molecular cloud with ice coating (`Dust2'), and for the surface of the disk associated with HH 30 (`Dust3'), respectively
and were used in \citet{Whitney03b,Whitney03a,Whitney04,Whitney13}.
{
Each dust model uses a number of homogeneous spherical particles with ``astronomical silicates" \citep{Draine84} and graphite, with the size distributions adjusted to reproduce some observations.
\citet{Whitney03b,Whitney03a} calculated the optical parameters (dust opacities, scattering albedos, forward throwing parameters, and polarization for scattered light) for various wavelengths for these dust models based on the Mie theory and the geometrical optics algorithm \citep{Wood02b}.
}
See Paper I for more details on these dust models
{and how to calculate the distributions of total intensity (Stokes $I$) at MIR wavelengths from polarized intensity distributions in the $H$-band.}

The dust opacities for these models are tabulated in Table  \ref{tbl:opacity}.
For this paper, we used the Dust1 and Dust2 models for the extended disks as well as the extended envelopes, as the Dust3 model cannot explain the optical properties of all of the known circumstellar disks associated with young stars.

\begin{table*}
\caption{Dust opacity. \label{tbl:opacity}}
\begin{tabular}{lccccccc}
\tableline\tableline
\multicolumn{1}{c}{$\lambda$} & \multicolumn{3}{c}{$\kappa_\mathrm{ext} (\lambda)$~~~(cm$^2$ $g^{-1}$)} 	&  \multicolumn{3}{c}{$\kappa_\mathrm{ext} (\lambda)$/$\kappa_\mathrm{ext} (H)$} \\
\multicolumn{1}{c}{(\micron)} & Dust1	& Dust2	& Dust3	& Dust1	& Dust2	& Dust3
\\ \tableline
1.65	($H$) & 37.5	& 60.8	& 52.4 	& 1		& 1 		& 1\\
3.81	& 8.30	& 15.3	& 18.1	& 0.22	& 0.25	& 0.35	\\
4.79	& 5.53	& 10.3	& 12.7	& 0.15	& 0.17	& 0.24	\\
11.33	& 8.29	& 11.1	& 6.67	& 0.22	& 0.18	& 0.13	\\
\tableline
\end{tabular} \\
\end{table*}

\begin{figure}[ht!]
\centering
\includegraphics[width=9cm]{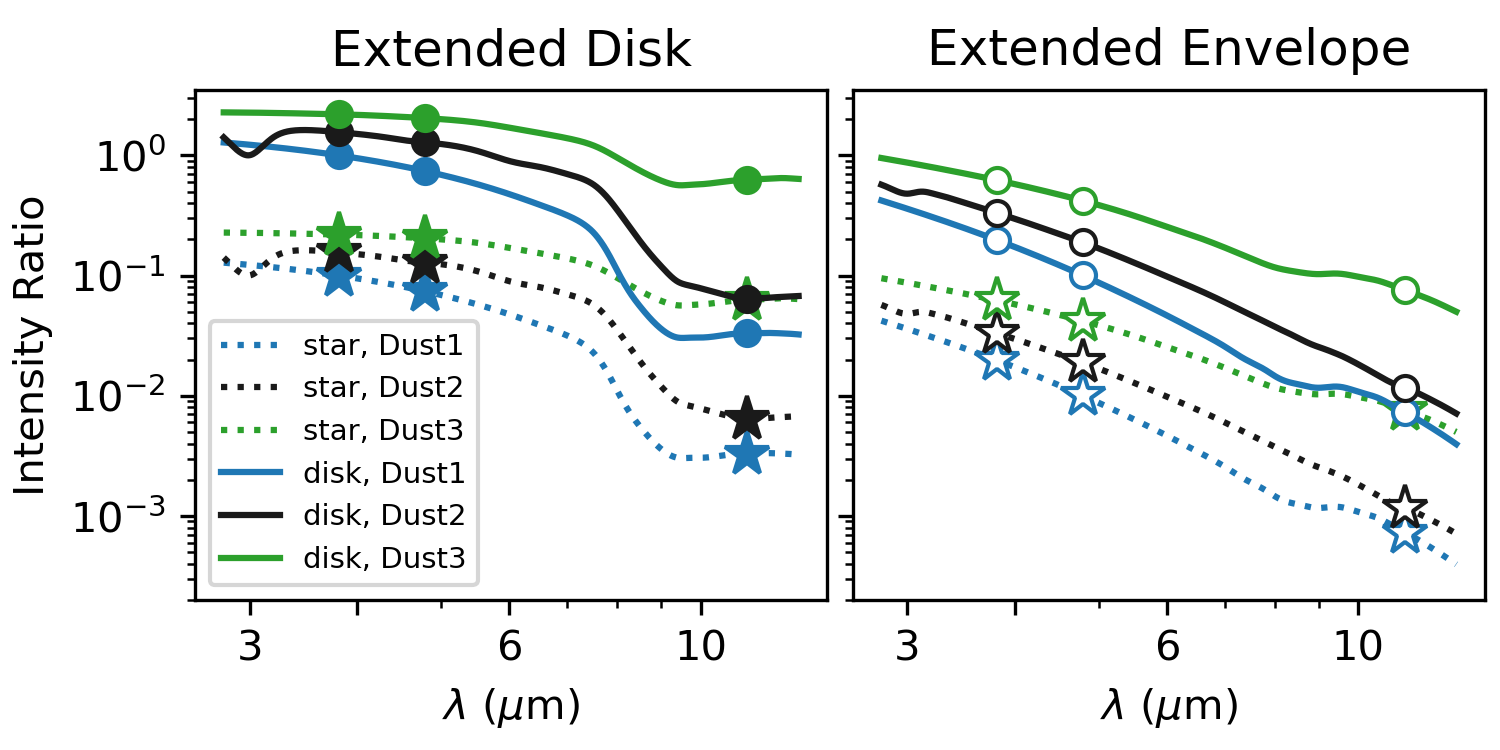}
\caption{
Relative intensities for the scattering emission for various cases. 
The left and right panels are for when we assumed that the extended emission is associated with a disk and an envelope, respectively.
The solid and dashed curves are for when the central illuminating source at NIR wavelengths is a self-luminous compact disk and a star, respectively.
The curves with different colors are for different dust models.
The dots indicate values at three representative wavelengths for the METIS observations ($\lambda$=3.8, 4.8, and 11.3 \micron).
All values are normalized to the intensity for an extended disk at $\lambda$=3.8 \micron~derived using a flat compact disk as the illuminating source at NIR wavelengths, and using the Dust1 model for the extended disk.
\label{fig:paper1_f15_revised}
}
\end{figure}

The calculated
{distributions of total intensity}
for the three MIR wavelengths ($\lambda$=3.8, 4.8, and 11.3 \micron)
{are}
dominated by the single scattering process, as was the case for those executed at similar wavelengths in Paper I. Figure \ref{fig:paper1_f15_revised} shows the calculated intensities for the single scattered light normalized by the following case: when both the bright central emission and the extended emission are from a disk with `Dust1'. The scattered emission is brighter for the following cases: (1) observations at short wavelengths; (2) when the extended emission is due to a disk rather than an envelope; (3) when the central radiation source is also a disk, not a star at any wavelength; and (4) for `Dust2' rather than `Dust1', and `Dust3' rather than `Dust2'.  Double scattering in the disk enhances the emission by up to $\sim$20 \% only.  The thermal emission is responsible for up to $\sim$10 \% 
at $\lambda$=11.3 \micron~in the regions close to the central source, but is negligible for all the other cases. 

When the central radiation source at NIR wavelengths is a star, the calculated intensities are also affected by the assumed typical grazing angle of the disk surface or the envelope $\overline{\gamma}$ ($\sim \overline{z/r}$, where $z$ and $r$ are the height and radius, respectively). 
In contrast, the intensity for single scattering is independent of $\overline{\gamma}$  if the central radiation source is a disk at all wavelengths. Therefore, the calculated intensity including all the radiation processes is nearly independent of $\overline{\gamma}$ for the latter case. The intensity derived using a star is smaller than that using a disk by approximately a factor of $\overline{\gamma}$. For this paper, we used $\overline{\gamma}$=0.1 as a lower limit to yield self-consistent calculations. See Paper I for details about the effects and limitations of $\overline{\gamma}$.


\subsection{Central source for HCI calculations} \label{sec:method:cs}

As described in Section \ref{sec:method:extended}, the bright central source is a self-luminous compact disk at the wavelengths of the HCI observations. Although this is nearly a point source at the given angular resolution, the emission from the outer disk radii leaks out from the coronagraphic mask, and therefore enhances the speckle noise, as demonstrated in later sections.

For this work, we used the following disk models: (1) conventional optically-thick accretion disk models \citep{Pringle81}; and (2) a Gaussian distribution with a HWHM of 1, 2, and 4 au.
Interferometric observations of disks at NIR and MIR wavelengths have been powerful in constraining their spatial distributions, but have not yet been able to determine detailed distributions as described below.
For (1), the intensity distribution depends on the wavelengths of the observations and the mass accretion rate (and therefore the resultant luminosity). This model successfully explained the observed SEDs and MIR interferometric visibilities ($\lambda \sim 10$ \micron) for FU Ori \citep{Quanz06}. This study was corroborated by \citet{Labdon21}, who executed multi-band NIR interferometry ($\lambda$ = 1.2-2.2 \micron) for the same star.
The models in (2) were used for the following studies. 
\citet{Liu19b,Lykou22} attributed their NIR-to-MIR interferometric observations ($\lambda$ = 2-3.5 \micron) of FU Ori to a Gaussian disk with a HWHM of 0.5-3 au.
\citet{Quanz06} demonstrated that the MIR interferometric visibilities ($\lambda \sim 10$ \micron) measured for FU Ori are consistent with a Gaussian disk with a HWHM of $\sim$4 au as well as the conventional disk model.

\begin{figure*}[ht!]
\centering
\includegraphics[width=18cm]{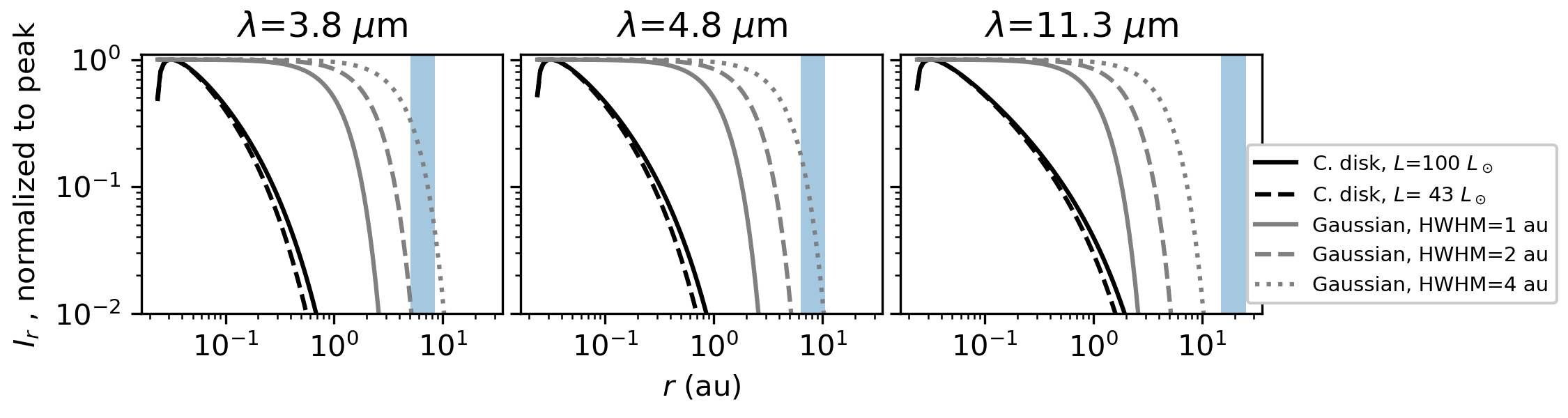}
\caption{
Radial intensity distributions for various disk models.
The left to right panels are for the three representative wavelengths for our calculations. 
Each panel shows the distributions for the conventional optically-thick accretion models (`C. disk') for the luminosities for FU Ori and V1735 Cyg (100 and 43 $L_\sun$, respectively; see Table \ref{tbl:targets}) and the Gaussian disk models for HWHM=1, 2, and 4 au. Each curve is normalized to the peak value.
The blue horizontal bar in each panel indicate the HWHM of the diffraction-limited PSF of the telescope at the target distances (Table \ref{tbl:targets}).
\label{fig:I_r_disk}
}
\end{figure*}

Figure \ref{fig:I_r_disk} shows the radial intensity distribution for the above disk models in three HEEPS bands.
The distribution of emission for the conventional disks was smaller than any of the Gaussian distributions we used. The conventional disks show a larger spatial extent at longer wavelengths because of contributions from the cooler outer regions. Their spatial extent is slightly larger for FU Ori than V1735 Cyg due to a larger disk luminosity. 
For all the cases, most of the flux from the central disk is distributed within the diffraction core of the telescope. 
As a result, their spatial distributions are almost identical to those of a point source without a coronagraph.

The inclination angle of the disk+envelope system is not known for our target objects (Paper I). For our simulation, we use these compact disks with a face-on view, which yields the maximum leakage from the coronagraphic mask and therefore conservative detection limits.

\subsection{HEEPS simulations} \label{sec:method:heeps}

To investigate the signal-to-noise ratio for the extended MIR emission, we used HEEPS v1.0.0\footnote{https://github.com/vortex-exoplanet/HEEPS} \citep{Delacroix22}, an open-source python-based software for HCI \citep{heeps22}.
HEEPS calculates the speckle noise associated with bright sources in the given detector format through the following steps:
(1)
derivation of a temporal series of single-conjugate adaptive optics (SCAO) residual phase screens, including the predicted instantaneous pointing errors of the observations;
(2)
propagation of the SCAO residual phase screens through the individual optical components of the instrument, using one of the vortex coronagraphy modes (CVC or RAVC);
(3)
accumulation of the instantaneous coronagraphic PSFs to produce a mock observing sequence in pupil-stabilised mode, 
including the METIS radiometric budget for the considered target star in the considered filter;
and
(4) computation of the post-processed contrast.
For each filter, these calculations are made monochromatically for the given representative wavelength.

Using HEEPS, we first calculated the following sequences (cubes) of the PSFs for the default field of view (FoV) of HEEPS of 2.2$\times$2.2 arcseconds:
\begin{description}

\item{(I)}
On-axis PSF cubes for the central sources described in Section \ref{sec:method:cs}. 
We calculated the angular distribution of the emission, and derived the PSF cubes for an {on-source} integration of 30 minutes for each wavelength, coronagraph (CVC or RAVC), and disk model. Each cube consists of 6000 images for the given integration and time interval.
\\ HEEPS has a built-in function to execute such simulations for slightly extended emission such as these disks, approximating them using multiple point sources. However, this function requires significant computational time proportional to the number of the point sources. To overcome this problem, we developed Monte-Carlo simulators to eject photon packets from the individual locations of the above disk models. We added 
the coordinate of each photon packet to the pointing error for the simulation for each image of the cube
provided with a 300-msec interval.
This method allowed us to calculate a PSF cube for each case with a computational time of approximately 45 minutes using an 8-12 CPU server.
The use of 6000 photon packets for a 30-minute integration reduced the statistical error for the intensity distribution of the PSF to within 2 \%. 

\item{(II)}
Off-axis PSF cubes used to convolve extended MIR emission derived in Section \ref{sec:method:extended}.
We also used the off-axis PSF at $\lambda$=3.8 \micron~to study the impact of a binary companion (Section 3.3).

\item{(III)}
On-axis PSF cubes for reference stars (point sources) to subtract the bright central emission from the target exposures (see Appendix \ref{app:no_psfsub} for necessity).
For these cubes, we used different sets of phase screens from (I)(II) in order to reproduce the observations at different times.
We used a star 3 times as bright as each target object, and executed simulations for a 30-minute integration which yielded 6000 frames. 
In Section \ref{sec:summary}, we will briefly discuss the use of stars with different brightness levels at varying integrations. \\
As described below, we derived median PSFs using the PSF cubes and subtracted them from the individual science frames. 
In the future, we may use PSF models with the Locally Optimized Combination of Images \citep[LOCI;][]{Lafreniere07} and Principal Component Analysis \citep[PCA;][]{Soummer12,Amara12,Juillard24} that have previously been successful when the central science target is a star.

\end{description}
%

{To calculate speckle noise, w}e used the {SCAO residual}
phase screens derived with a 300 msec sampling for the following conditions{,} as in \citet{Carlomagno20,Delacroix22}.
{We split the currently available 12000 phase screens into two to obtain the cubes for (I)(II) and (III), respectively.}
The mock observations were made for median atmospheric conditions at the ELT site, at an assumed declination of $-$5\arcdeg~and with a $K$-band ($\lambda$=2.2 \micron) magnitude of 5 for SCAO corrections. The actual target $K$ magnitudes of 5-7 for FU Ori and V1735 Cyg yield similar performances as our mock observations \citep{Feldt24}.
The assumed declination yields a rotation of the parallactic angle of about 20$^\circ$ during the given integration when the target was crossing the meridian.
{For the thermal background, we used constant values estimated for the individual filters tabulated in Table \ref{tbl:inst}. We approximated that the PSF subtraction described above does not yield any residual patterns for the thermal background except for the photon noise.}

Using the cubes (I)-(III), one would execute the following simulations for extended emission in order to approximate the actual observations in the sky:
(A1) apply field rotations to the extended emission for the mock observing sequence;
(A2) convolve the images of the extended emission using the off-axis PSFs calculated for the individual time sequences;
(A3) add the on-axis PSF for the central source;
(A4) subtract the stacked reference PSF;
(A5) de-rotate the images to match their sky coordinates; and
(A6) stack the cube to create the final image.
However, this sequence also requires significant computational time due to the number of combinations of the bands, coronagraphic modes, central sources, and extended emission in our study. In practice, the process (A5) returns the location of the extended emission to its original position before applying the field rotations (A1). Therefore, we alternatively executed the following processes, which yielded identical results but with a significantly smaller computational time:
(B1) subtracted the stacked reference PSF from each image in the on-axis PSF cube;
(B2) derotated the images in the on-axis and off-axis PSFs, respectively, to match their sky coordinates;
(B3) stacked the on-axis and off-axis PSF cubes, respectively; and
(B4) convolved the extended emission using the stacked off-axis PSF and added it to the stacked on-axis PSF.

Before the PSF subtraction, we stacked the reference PSF without field rotation, and scaled it using the fluxes of the target PSF (without the extended emission) and the reference PSF measured within a radius of 0\farcs5. This scaling process yielded the cases when the ideal PSF subtraction was achieved. In contrast, the images obtained through actual observations of the targets would include both the on-axis PSF and extended emission, and therefore require complicated optimization of the PSF subtraction to avoid over-subtraction. This optimization is beyond the scope of this paper, that is, to investigate whether the target emission is actually detectable over speckle and photon noise.

The photon noise was added to the final images as follows. Before the process (B1), we first obtained the image of the photon counts for the stacked reference PSF with the thermal background, derived the image of the Poisson noise, and added it to the stacked reference PSF. Secondly, we obtained the image of the photon counts for the target frames with the thermal background, through (B2)-(B4), but without PSF subtraction. We then derived the distribution of the Poisson noise and added it to the image derived through (B1)-(B4). This method yields the same level of photon noise as we add Poisson noise to each image with 300-msec sampling, but with significantly fewer random numbers and therefore total time for calculation.

We then converted the units of the extended emission to W m$^{-2}$ \micron$^{-1}$ arcsec$^{-1}$.
In addition to the images with extended emission, we also created images using the same PSFs but without extended emission. We used these images to estimate the detection limits of the extended emission in later sections.

The read noise of the detector{, which is not included in our calculations,} is marginal or negligible compared with the total noise if we select a sufficiently long exposure. 
{Under these conditions, the results presented in the rest of the paper are independent of the actual exposure time.}
Such exposures {saturate} the central source in limited cases, but do not significantly affect the observations of the extended emission in which we are interested (Appendix \ref{app:saturation}).
{To minimize the overheads for the detector readouts, and therefore the total time for the observations, one would select an actual exposure time to be as long as possible without the detector saturation significantly hampering the image alignment for the image stacking process.}


\section{Results} \label{sec:results}

In Section \ref{sec:results:basic}, we summarize the basic coronagraphic performance for various cases without target extended emission.
In this subsection, we demonstrate how the different central compact disks (the conventional disks and the Gaussian disks with a HWHM of 1, 2, and 4 au) yield different PSFs and different distributions of speckle noise.
In Section 3.2, we present the simulations of the target extended emission.
In Section 3.3, we demonstrate how the existing binary companion of FU Ori affects the observations of the extended emission.

%
\subsection{Basic coronagraphic performances} \label{sec:results:basic}

\begin{figure*}[ht!]
\centering
\includegraphics[width=14cm]{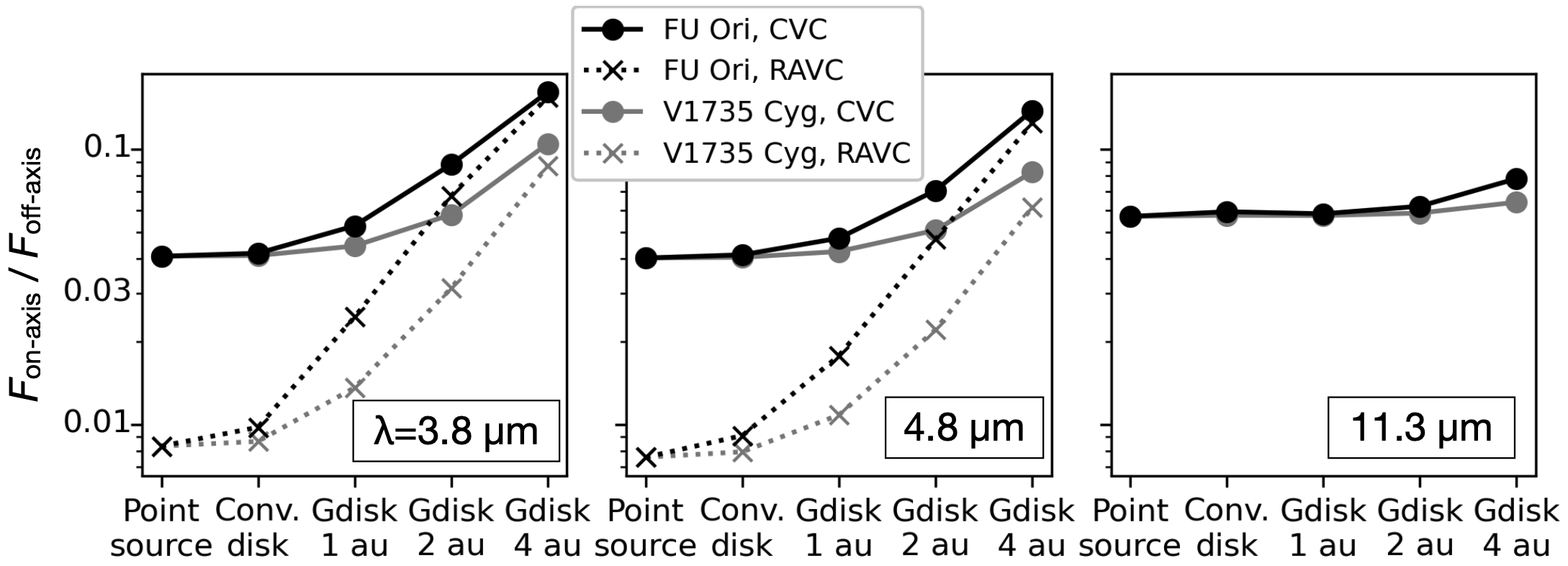}
\caption{
Reduction of the flux of the central disk by the coronagraphic mask for various cases. These are shown as the flux ratios for the on-axis and off-axis observations. The left to right panels show the ratios for the three wavelengths. Each panel shows the ratios when the central source is FU Ori and V1735 Cyg, and observed using two coronagraphs (CVC and RAVC) to be installed in HCI. The horizontal axis of each panel shows the cases where the central source is a point source, the conventional accretion disk, and Gaussian disks with a HWHM of 1, 2, and 4 au, respectively. These are organized from left to right for smaller to larger spatial distributions (Figure \ref{fig:I_r_disk}).
\label{fig:on_per_off}
}
\end{figure*}

Figure \ref{fig:on_per_off} shows how the two coronagraphs work to reduce the flux from the bright central disk in various cases, that is, for different disk models and three bands, and when the target is FU Ori or V1735 Cyg. These are shown as the flux ratios of the on-axis (when the emission is centered on the coronagraphic mask) to the off-axis observations (when the emission is out of the coronagraphic mask) measured within $r$=5$\lambda$/D of the central source (0\farcs12, 0\farcs15, and 0\farcs37 at $\lambda$=3.8, 4.8, and 11.3 \micron, respectively; where $D$ is the diameter of the telescope).
This region for each wavelength would cover 90--95 \% of the total flux without a coronagraph.
The derived ratios, which differ from those using the conventional method with the central position only \citep[``peak rejection rate"; e.g.,][]{Carlomagno20}, allow us to evaluate the flux reduction independent of different shapes of PSFs.

In each panel of Figure \ref{fig:on_per_off}, we also show the flux ratios for a point source for reference.
The horizontal axis of each panel is organized from left to right for small to large spatial extent of the bright central source
(Section \ref{sec:method:cs}; Figure \ref{fig:I_r_disk}).
Speckle noise and the pointing errors for the AO were included but photon noise is not.

In general, a large spatial extent for the bright central source yielded a larger flux leakage from the mask. However, the differences were marginal between the following cases: (1) the central source is a point source and with the conventional accretion disk for the observations at $\lambda$=3.8 and 4.8 \micron, and (2) all the cases for the observations at $\lambda$=11.3 \micron.
The latter trend is attributed to the relatively large diffraction pattern at the wavelength of the observations (Figure \ref{fig:I_r_disk}). The coronagraph worked better for V1735 Cyg than FU Ori, in particular for the Gaussian disk models, because of the smaller angular scales at a larger distance. 
At $\lambda$=3.8 and 4.8 \micron, for which both of the CVC and RAVC coronagraphs are available, RAVC yielded a better capability to reduce the flux of the central source by a factor of up to $\sim$5.

\begin{figure*}[ht!]
\centering
\includegraphics[width=17cm]{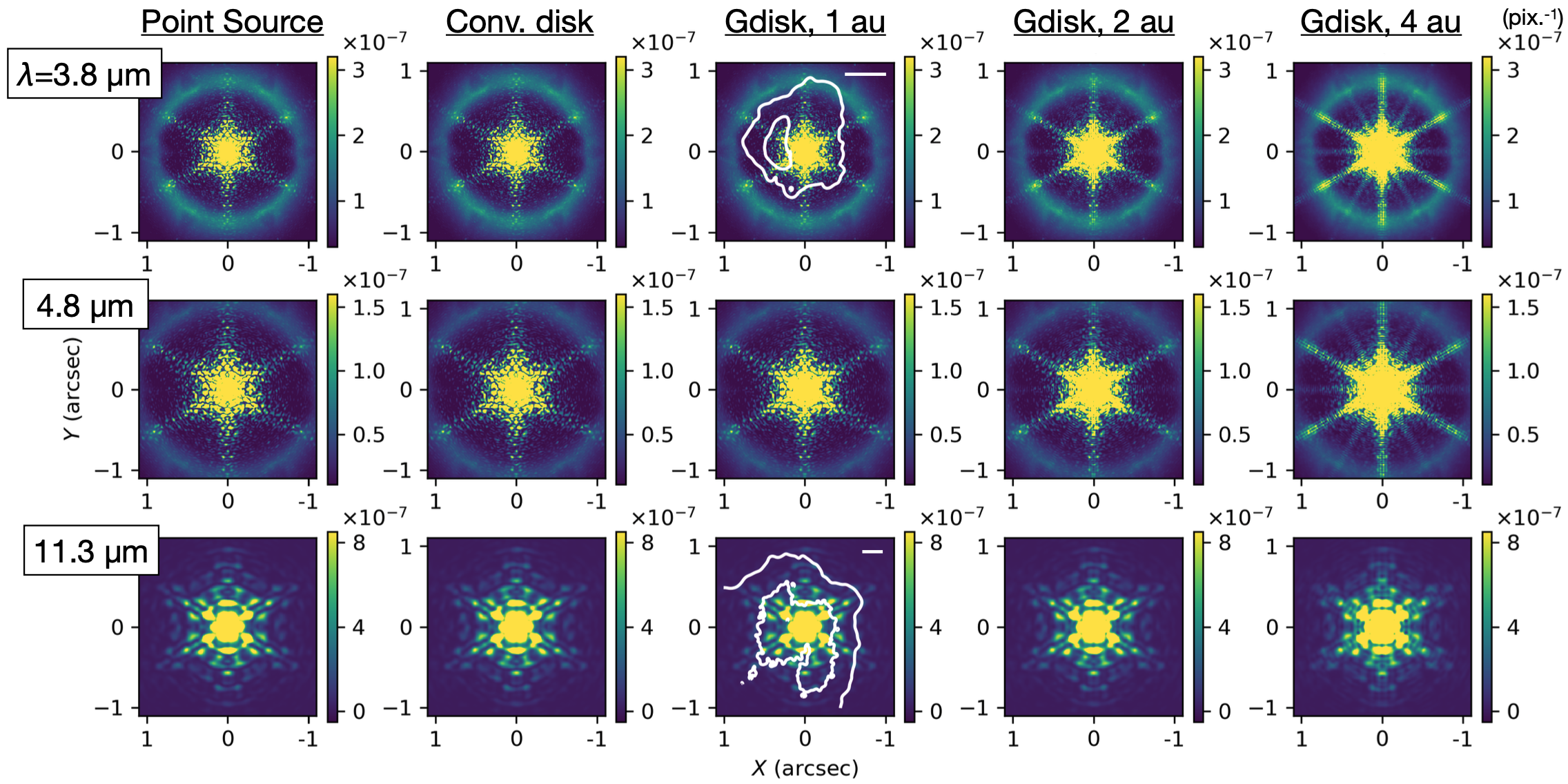}
\caption{
Simulated scenes for FU Ori, with marginally resolved inner disks,
with the CVC coronagraph for various cases.
These scenes do not include any emission from an extended disk or envelope as described in Section \ref{sec:method:extended}.
The panels from left to right are those where the central source is a point source, the conventional accretion disk, and the Gaussian disks with a HWHM of 1, 2, and 4 au, respectively, i.e., for small to large sources. The panels from top to bottom are for $\lambda$=3.8, 4.8, and 11.3 \micron, respectively.  The intensity at each pixel is normalized by the total off-axis transmission.
The images for the whole integration were stacked in the detector coordinates without field rotation.
For reference, we show some contours in the middle panels to indicate an approximate spatial extension of the extended emission associated with FU Ori (the upper middle panel) and V1735 Cyg (the lower middle panel) shown in Figure \ref{fig:PI_H}.
The horizontal bars at the top-right of these panels indicate a spatial scale of 200 au.
\label{fig:psfs_cvc}
}
\end{figure*}

Figure \ref{fig:psfs_cvc} shows the on-axis PSFs for FU Ori with the CVC coronagraph for three bands and different central disks.
To show the instrumental PSFs, we stacked the PSF cubes without field rotation.
At $\lambda$=3.8 and 4.8 \micron,
the PSF for each central source consists of a bright core with six diffraction spikes at 60$^\circ$ intervals, and a ring with a diameter of about 1 arcsecond. 
The six diffraction spikes are due to 
the shadow of the support structure that hold the secondary mirror of ELT.
The ring corresponds to the {control radius} of the SCAO with the ELT-M4 deformable mirror {(0.8, 1.0, and 2.5 arcsec for $\lambda$=3.8, 4.8, and 11.3 $\micron$, respectively)}.

We find marginal differences in the PSFs between the smallest and largest disks shown in the left and right ends of the figure. When the central source is larger, the core and the spikes are brighter while the brightness of the ring remains the same. The brightest PSFs at $\lambda$=3.8 and 4.8 \micron~are also associated with six more faint diffraction spikes between the six bright spikes.
The PSFs at $\lambda$=11.3 \micron~resemble the central part of the PSFs at $\lambda$=3.8 and 4.8 \micron.
The PSFs for the RAVC coronagraph also show the same trends described above (see Appendix \ref{app:psf_ravc}).

\begin{figure*}[ht!]
\centering
\includegraphics[width=18cm]{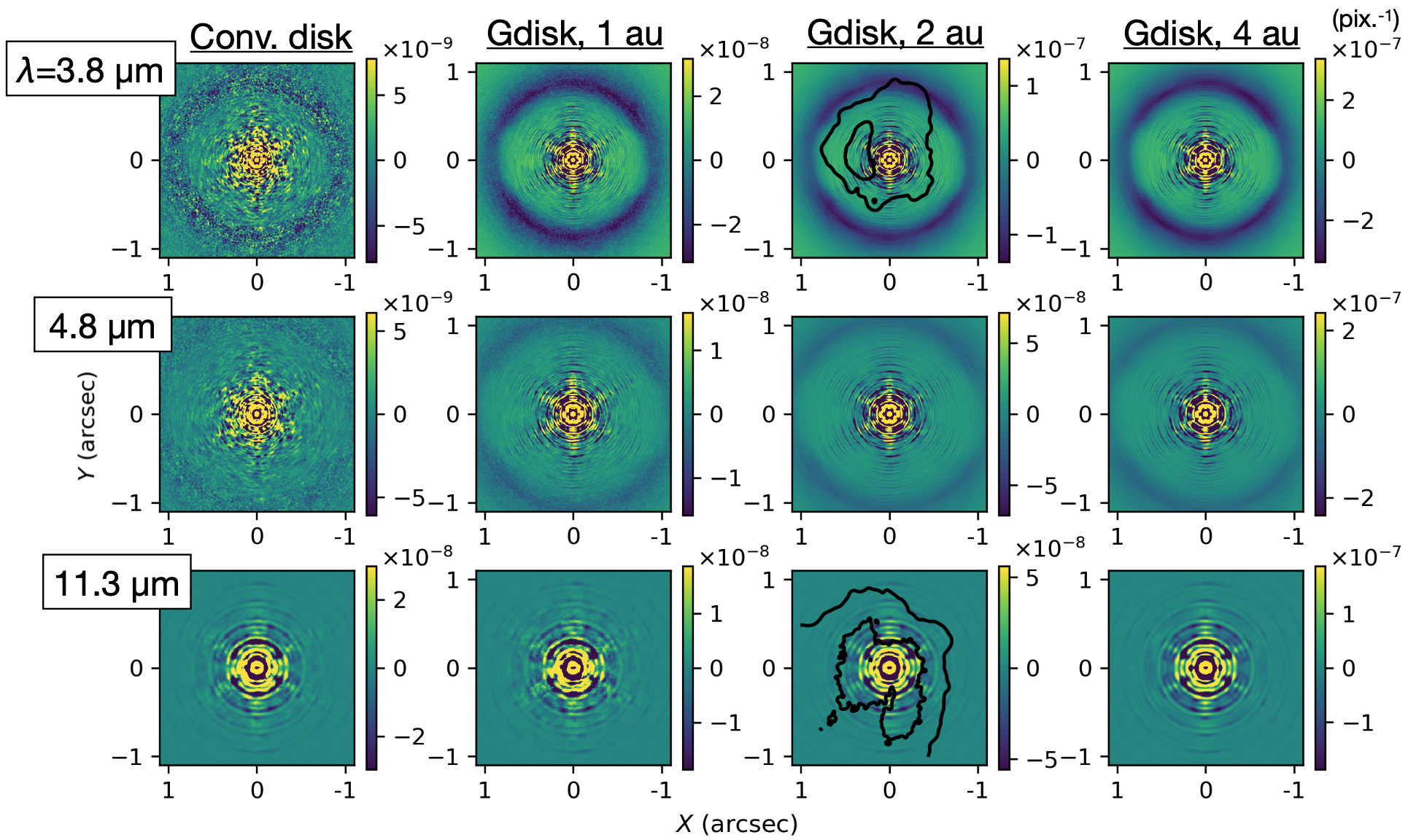}
\caption{
Same as Figure \ref{fig:psfs_cvc} but for the four central disks after subtracting the reference PSF and applying the field derotation before stacking.
The color bar for each panel is adjusted to clearly show the residual intensity distribution in the central part.
\label{fig:psfsub_cvc}
}
\end{figure*}

Figure \ref{fig:psfsub_cvc} shows the same PSFs but after subtracting the reference PSF and applying the field rotation before stacking (B2 and B3 in Section \ref{sec:method:heeps}) in order to show how the residual of PSF subtraction can affect the observations of the extended emission.
The residual pattern of the PSF subtraction does not show significant differences between different central disks but does in the absolute level of the speckle noise associated with the spike patterns, and a negative ring occurs due to the flux scaling of the reference PSF. The use of the RAVC coronagraph yields a similar trend (Appendix \ref{app:psf_ravc}).
For both cases, the spike patterns become significantly less noticeable because of (1) the PSF subtraction; and (2) the field derotation of the individual frames before stacking.
For FU Ori and V1735 Cyg, the major features of our interest in the extended emission are located within the negative ring pattern described above.

\begin{figure*}[ht!]
\centering
\includegraphics[width=14cm]{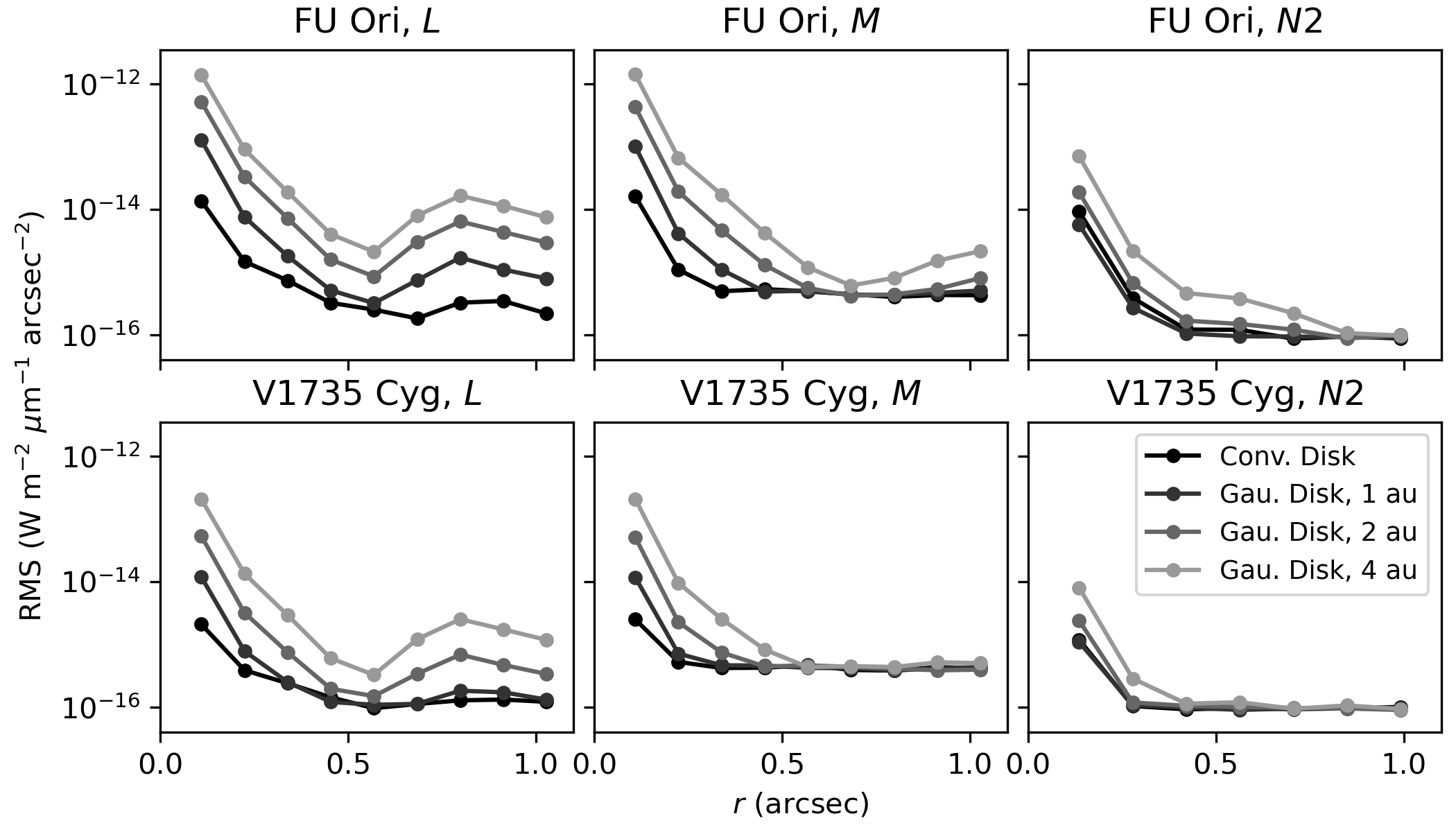}
\caption{
The {5}-$\sigma$ detection limits for the CVC observations. The top and bottom panels are for different targets, and the left to right panels are for observations at short to long wavelengths.
For each panel, we show the detection limits as a function of distance from the central source, which is either the conventional accretion disk or a Gaussian disk with HWHM=1, 2, or 4 au.
The noise was measured along the horizontal axis (i.e., between the spike patterns that cause larger noise levels) after convolving the images using a Gaussian function with a FHWM of 30 mas.
\label{fig:r_vs_noise_cvc}
}
\end{figure*}

Figure \ref{fig:r_vs_noise_cvc} shows the {5}-$\sigma$ detection limits of the extended emission for various cases of the CVC observations as a function of angular distance from the central disk. The calculation for each dot was made by measuring the root mean square of the speckle and thermal noise in a 21$\times$21-pixel box (approximately 0\farcs1$\times$0\farcs1; see Table \ref{tbl:inst}) between the residual of the bright spikes after the PSF subtraction. Before the measurements, we convolved the image using a 2-D Gaussian with a FWHM of 30 mas to improve the detection limit without significantly degrading the images for the target extended emission (Section 3.2).

At longer wavelengths ($\lambda$=4.8 and 11.3 \micron), the constant noise level in the outer radii is due to photon noise. Any other curves, whose spatial variations are due to the speckle noise, show that the detection limits are smaller for larger distances at the inner radii, but these increase at the outer radii due to speckle noise associated with the negative rings (Figure \ref{fig:psfsub_cvc}).

\begin{figure}[ht!]
\centering
\includegraphics[width=9cm]{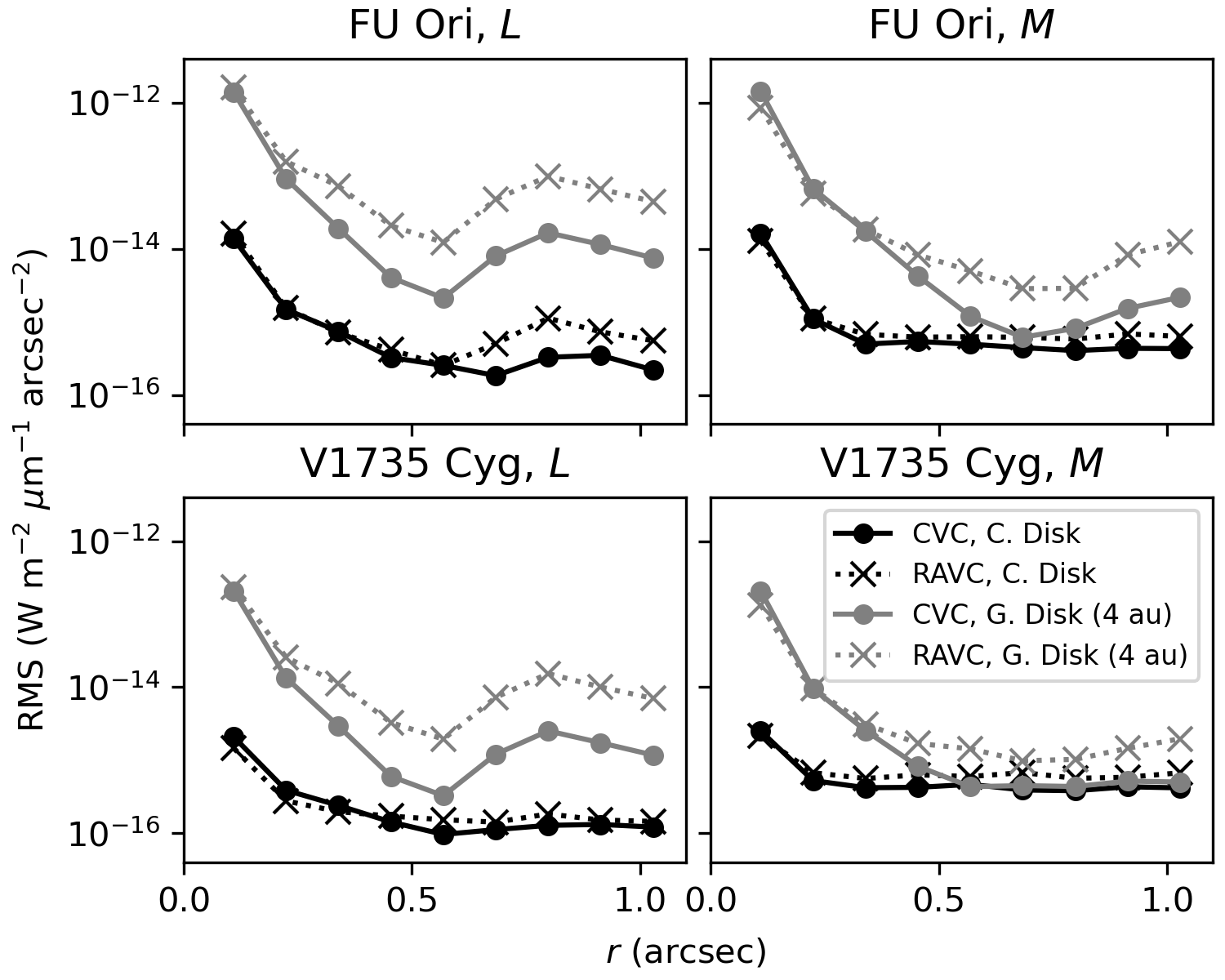}
\caption{
Same as Figure \ref{fig:r_vs_noise_cvc} but for both of the CVC and RAVC coronagraphs. The {5}-$\sigma$ detection limits shown in the figure are limited to the smallest and largest disks and for $\lambda$=3.8 and 4.8 \micron~in order to clarify the differences between the two coronagraphic modes.
\label{fig:r_vs_noise_cvc_ravc}
}
\end{figure}

In Figure \ref{fig:r_vs_noise_cvc_ravc}, we compare the detection limits for the CVC and RAVC coronagraphs using the smallest and largest central disks. CVC yields better detection limits than RAVC.
Therefore, we use CVC for the rest of the paper in order to investigate the detection of the target extended emission for various cases.

%
\subsection{Observations of the target extended emission} \label{sec:results:extended}

\begin{figure*}[ht!]
\centering
\includegraphics[width=18.4cm]{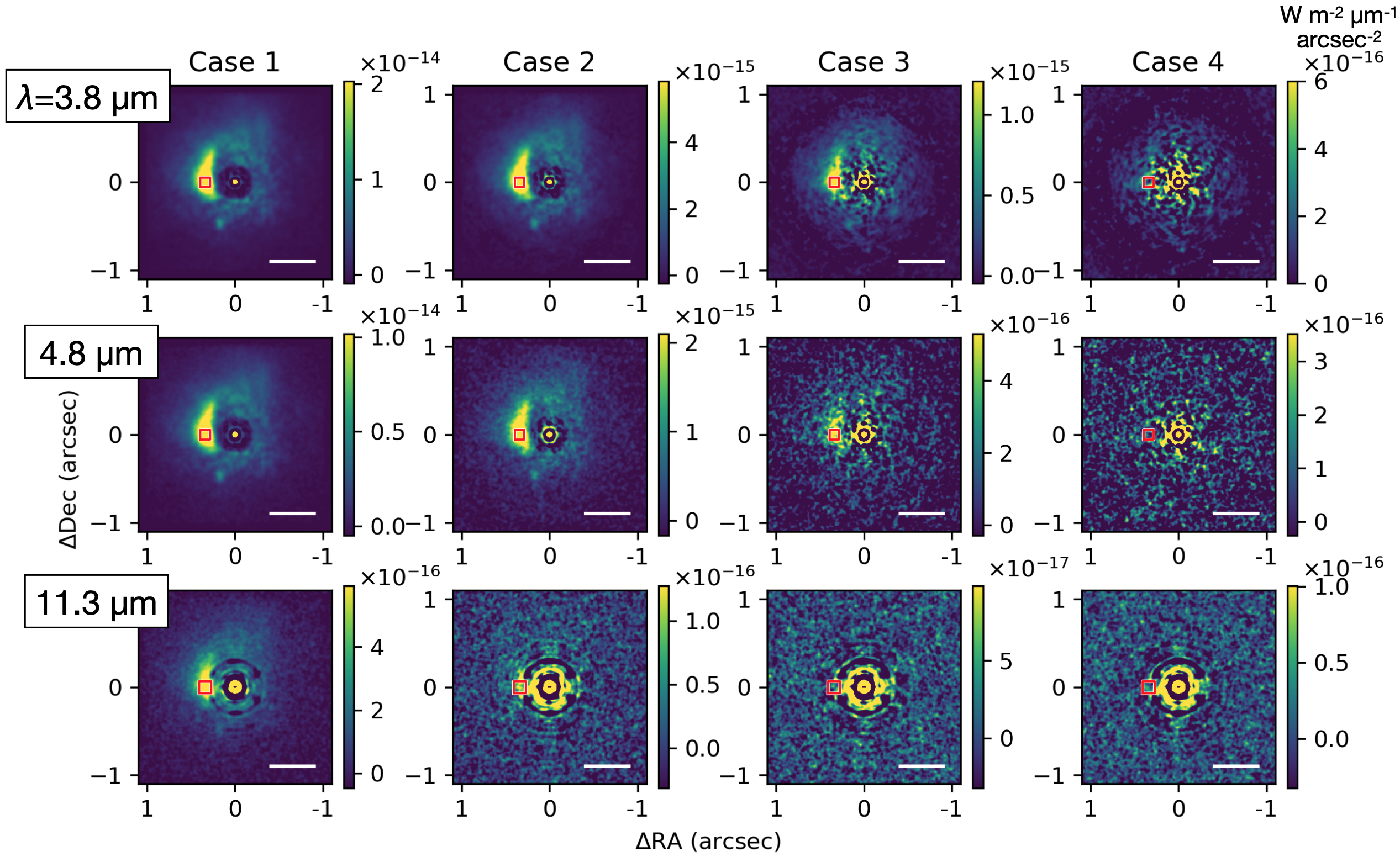}
\caption{
Simulated images for FU Ori, with MIR emission from an extended disk/envelope,
using the CVC coronagraph. The panels from left to right are for Cases 1 to 4, as tabulated in Table \ref{tbl:cases}. The panels from top to bottom are for $\lambda$=3.8, 4.8, and 11.3 \micron,
respectively. The central source in the MIR is the conventional accretion disk.
The PSF reference is scaled and subtracted.
Each image is convolved with a Gaussian with a HWHM of 30 mas to increase the signal-to-noise of the target extended emission.
The small red square in each image shows the position where we measured the signal-to-noise for Figure \ref{fig:sn}.
The horizontal bar at the bottom-right of each panel indicates a spatial scale of 200 au.
\label{fig:FU_4}
}
\end{figure*}

\begin{figure*}[ht!]
\centering
\includegraphics[width=18.4cm]{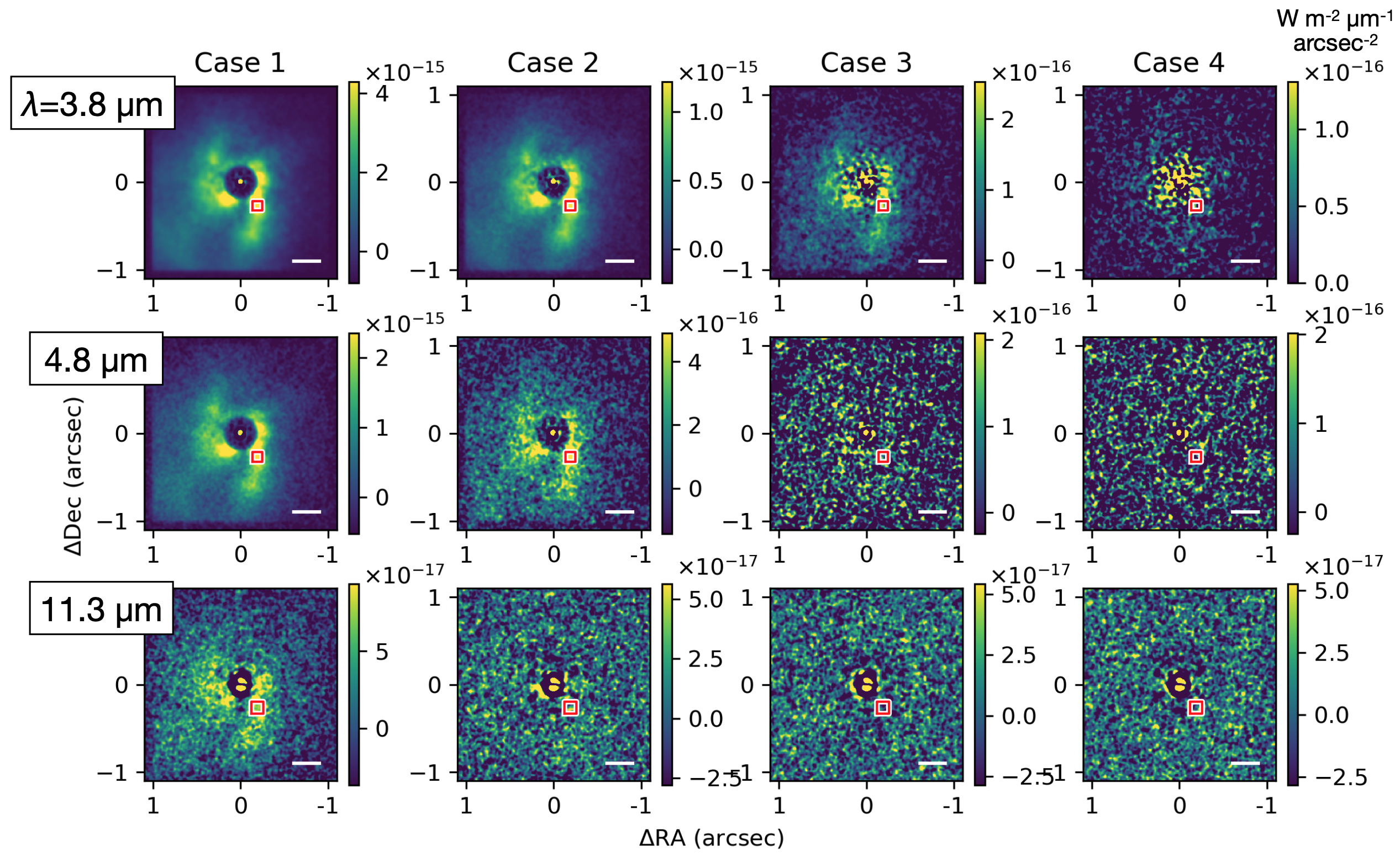}
\caption{
Same as Figure \ref{fig:FU_4} but for V1735 Cyg.
\label{fig:V17_4}
}
\end{figure*}

Figures \ref{fig:FU_4} and \ref{fig:V17_4} show PSF-subtracted images for four combinations of the extended emission (a disk or an envelope), the central radiation source at NIR wavelengths (a disk or a star), and dust grain models (Dust1, 2, and 3). These combinations are tabulated in Table \ref{tbl:cases} as Cases 1-4. Cases 1 and 2 are the brightest cases where we use the extended disk and the envelope, respectively. Cases 3 and 4 are the same but for the faintest cases (see Figure \ref{fig:paper1_f15_revised}). For all cases, we used the conventional accretion disk as the bright central source in the MIR. For reference, we also measured the signal-to-noise at the peak positions indicated in the figures. These values are tabulated in Table \ref{tbl:sn}.

In general, the observations at shorter wavelengths yielded better signal-to-noise ratios due to the brighter nature of the target extended emission (Figure \ref{fig:paper1_f15_revised}) and significantly fainter thermal background (Table \ref{tbl:inst}).
In contrast, the target extended emission can be observed only for the brightest cases at $\lambda$=11.3 \micron.

\begin{table}
\caption{Signal-to-noise at the peak position \label{tbl:sn}}
\hspace{-1.8cm}
\begin{tabular}{lccccc}
\tableline\tableline
Target	& $\lambda$ (\micron)	& Case 1	& Case 2 	& Case 3	& Case 4
\\ \tableline
FU Ori		& 3.8		& 2.3$\times 10^2$	& 65		& 11	 	& 2.1 \\
			& 4.8		& 1.5$\times 10^2$	& 31		& 5.5		& 0.8 \\
			& 11.3	& 17				& 2.0	 	& 0.09	& 0.02 \\
V1735 Cyg	& 3.8		& 99				& 28		& 4.5		& 0.9 \\
			& 4.8		& 33				& 6.8		& 1.2		& 0.2 \\
			& 11.3	& 4.8				& 0.6		& 0.03	& 6$\times10^{-3}$ \\
\tableline
\end{tabular}
\end{table}


\begin{table}
\caption{Parameters of four cases for extended emission. \label{tbl:cases}}
\begin{tabular}{llcl}
\tableline\tableline
Case		& Extended 	& Central Radiation 	& Dust \\
		& Emission	& Source \\
		&			& @NIR
\\ \tableline
1	& Disk		& Disk	& Dust3 \\
2	& Envelope	& Disk	& Dust3 \\
3	& Disk		& Star	& Dust1\\
4	& Envelope	& Star	& Dust1\\
\tableline
\end{tabular} \\
\end{table}

\begin{figure*}[ht!]
\centering
\includegraphics[width=18.4cm]{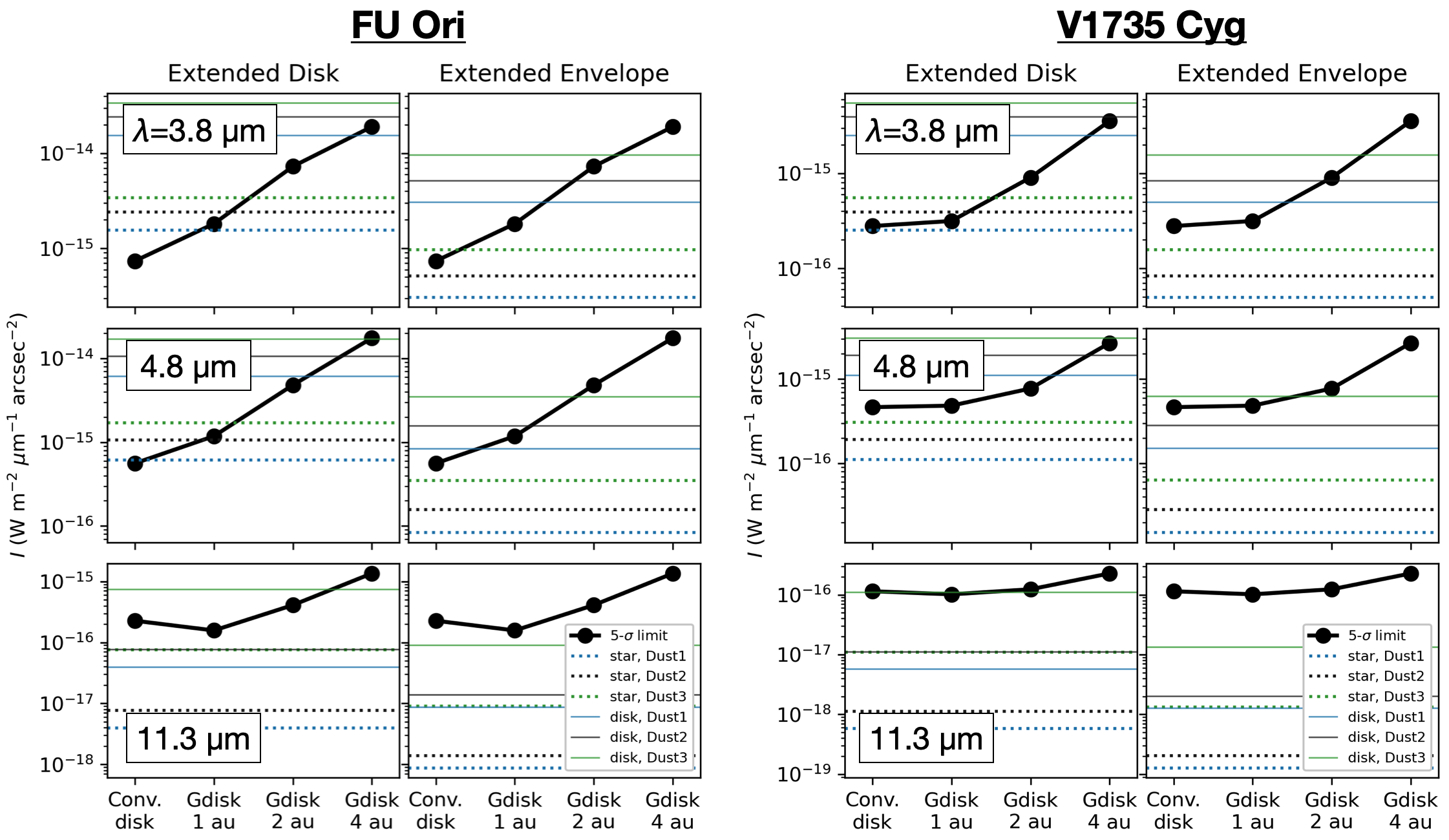}
\caption{
Comparisons between the {5}-$\sigma$ detection limits (black lines with dots) and the brightness of the extended emission (horizontal lines).
The left and right panels are for FU Ori and V1735 Cyg, respectively. 
The top to bottom panels are for $\lambda$=3.8, 4.8, and 11.3 \micron, respectively.
For each target and band, we organize two panels to show the values when the extended emission is associated with a disk ($left$) and an envelope ($right$).
In each panel, we plot the brightnesses for different central radiation sources at NIR wavelengths (a star or a disk) and for different dust models (Dust1, 2, and 3). The left to right dots in each panel are the values when the bright central source in the MIR is the conventional accretion disk and Gaussian disks at HWHM of 1, 2, and 4 au.
The {5}-$\sigma$ detection limits for the individual cases were measured at the positions shown in Figures \ref{fig:FU_4} and \ref{fig:V17_4}.
\label{fig:sn}
}
\end{figure*}

The use of the Gaussian disks instead of the conventional accretion disk yields similar images to those in Figures \ref{fig:FU_4} and \ref{fig:V17_4},
(but with comparable or lower signal-to-noise ratios) for the following reasons: (1) the major emission features shown in these figures are located within the negative ring pattern caused by the PSF subtraction; and (2) these central compact disks yield similar noise pattern but the absolute intensity scale within the ring (see Section \ref{sec:results:basic}). 
In Figure \ref{fig:sn}, we plot the brightness of the target extended emission and {5}-$\sigma$ detection limits at the emission peaks indicated in Figures \ref{fig:FU_4} and \ref{fig:V17_4}. If the bright central source in the MIR is the conventional accretion disk, the observations at $\lambda$=3.8 \micron~will allow us to detect extended emission over {5}-$\sigma$ for many cases ({8-10} out of 12 for each of FU Ori and V1735 Cyg). 
At $\lambda$=4.8 \micron, the detections will be limited to {4}-9 cases for each target.
At $\lambda$=11.3 \micron, the extended emission will be detected only for the brightest cases. These are: when the central radiation source at NIR wavelengths is a disk, and the extended emission is due to a disk with Dust3. 

The detection rates are worse for larger central disks. If it is a Gaussian disk with a HWHM of 4 au, the extended emission will be detected only at $\lambda$=3.8 and 4.8 \micron, if the central radiation source at NIR wavelengths is a disk, and the extended emission is associated with an extended disk.

%

\subsection{Simulations with a companion star} \label{sec:results:companion}

FU Ori is known to be associated with a binary companion (FU Ori S) at a separation of 0\farcs5 \citep{Wang04, Reipurth04}. \citet{Reipurth04} measured $L'$-band ($\lambda = 3.8$ \micron) magnitudes of FU Ori and FU Ori S of 4.2 and 8.1, respectively. This companion was only marginally visible in the images in Section 3.2 because the imaging polarimetry technique used for the original $H$-band images significantly reduced its flux (see Figure \ref{fig:PI_H}).

\begin{figure*}[ht!]
\centering
\includegraphics[width=15cm]{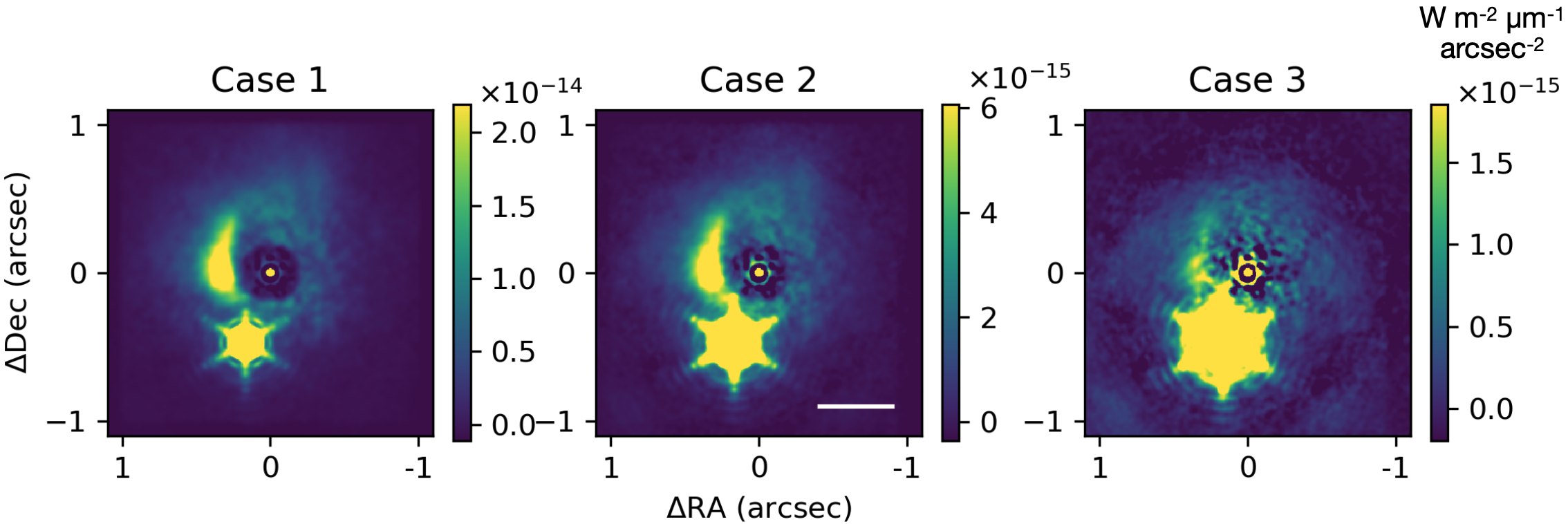}
\caption{
Simulated images of FU Ori with FU Ori S at $\lambda$=3.8 \micron~using the CVC coronagraph. The panels from left to right are for Cases 1-3 (see Table \ref{tbl:cases} for their parameter sets). The bright central source at the observing wavelength is assumed to be the conventional accretion disk. The color contrast for each image is arbitrarily adjusted to investigate how the observations of the extended emission suffer from the companion.
The horizontal bar at the bottom-right of the middle panel indicates a spatial scale of 200 au.
\label{fig:companion}
}
\end{figure*}

Figure \ref{fig:companion} shows the simulated images with the binary companion and the extended emission at $\lambda$=3.8 \micron~with the CVC coronagraph and for Cases 1-3. As in Figure \ref{fig:FU_4}, the bright central source at the observing wavelength is assumed to be the conventional disk. 
As shown in Figure \ref{fig:FU_4}, these are the best cases for increasing the signal-to-noise ratio for the extended emission in terms of the observing wavelength (see Figure \ref{fig:FU_4}) and the bright central source in the MIR (Figure \ref{fig:sn}). Figure \ref{fig:companion} still shows the bright arm in the east. However, these images suggest that the presence of a binary companion can significantly hamper observations of the extended emission if they spatially overlap.

This problem could be resolved if we use a PSF reference for the binary companion as well as the bright central emission. We need different integrations for these PSFs than those for the bright central source because the companion and the bright emission are located off-axis and on-axis of the coronagraph optics, respectively.


\section{Conclusions and discussion} \label{sec:summary}

METIS-HCI observations of the extended emission associated with FUors have great potential to yield a breakthrough in understanding their nature, and therefore further details of their stellar accretion and possible planet formation scenarios. This understanding would be applicable to many YSOs as well. According to our analysis with a limited sample (FU Ori, V1735 Cyg), the detection of infrared extended emission using HCI can be affected by many uncertainties. These include: the central radiation source at NIR wavelengths (either a self-luminous disk or a star); the spatial extent of the bright central disk emission in the MIR; whether the extended emission is associated with a disk or an envelope; and the nature of the internal dust grains responsible for the infrared emission. 

We investigated the detectability of the extended emission for the three representative METIS bands ($\lambda$=3.8, 4.8, and 11.3 \micron) and the two coronagraphic modes (CVC/RAVC) that will offer a field of view sufficient for the target extended emission.
Of the three wavelengths, the observations at $\lambda$=3.8-\micron~will yield the highest chance of detecting the extended emission due to its bright nature and relatively low thermal background. The observations at this wavelength will also have the advantages described below. First, these will improve the angular resolution by a factor of $\sim$2 compared with the existing NIR observations {(mainly in the $H$-band at $\lambda$=1.65 $\micron$)} at 8-m class telescopes, hopefully allowing us to better investigate the origin and the details of the observed structures. Secondly, the observations at short wavelengths suffer less from the uncertainties in the intensity of the extended emission (Paper I; Figure \ref{fig:paper1_f15_revised}).
CVC yielded better detection limits than RAVC.

In general, the dust opacity is smaller at longer wavelengths (Table \ref{tbl:opacity}).
That at  $\lambda$=3.8 \micron~would be 3-5 times lower than the $H$-band ($\lambda$=1.65 \micron).
If the 3.8-\micron~image shows a spatial distribution different from the $H$-band, it suggests that the extended envelope contributes to the $H$-band emission, while the 3.8-\micron~emission is associated with regions closer to the surface of the extended disk \citep[or hopefully the disk surface; Section \ref{sec:intro};][]{Fukagawa04,Hashimoto11}. In this case, a typical (3.8-\micron)/H
intensity ratio observed in the extended emission (hereafter $I_\mathrm{3.8\micron}/I_\mathrm{H}$)
would allow us to estimate the brightnesses at longer MIR wavelengths such as $\lambda$=4.8 and 11.3 \micron, therefore allowing us to investigate whether we will be able to detect emission at these wavelengths to observe the circumstellar structures even closer to the disk surface.
If the images at $\lambda$=3.8-\micron~and $H$-bands show the same spatial distribution, the extended emission at these wavelengths must share the same origin.
In this case, the $I_\mathrm{3.8\micron}/I_\mathrm{H}$ ratio will allow us to investigate whether the extended emission is associated with a disk or an envelope (Paper I). As such, we would be able to identify the origins of the observed structures and discuss the implications for star and planet formation. 
One may start the observations from FU Ori, which is used for the analysis in this paper and is observable from ELT. Observations of even a single target would allow us to investigate the observability of the other FUors, assuming that they share the same origins for the central radiation source at NIR wavelengths, for the bright central disk emission at MIR wavelengths, and for the dust grains responsible for the extended infrared emission.

Alternatively, one may execute MIR imaging observations (as well as the NIR observations for the other FUors) using existing telescopes prior to the operation of the ELT. Such observations would also be useful for investigating the detectability of the extended emission prior to METIS observations. Preliminary imaging observations using existing telescopes would also be useful to investigate whether target FUors other than FU Ori are associated with a binary companion that potentially hampers studies of the MIR extended emission.

For any of the above MIR wavelengths, we need to subtract the bright central PSF using a reference PSF from a star in order to execute detailed analysis of the extended emission (Appendix \ref{app:no_psfsub}). Throughout our simulations, we used reference stars three times brighter than the target objects and applied the same integration time as the targets. The observing time for the reference PSF could be reduced, but we would need a brighter source as the photon noise for the reference frames is enhanced when we scale them before subtracting them from the science frames (Section \ref{sec:method:heeps}). 
Alternatively, new software techniques such as MAYONNAISE \citep{Pairet21} and REXPACO \citep{Flasseur21} would be useful for removing the PSF even without using the conventional PSF subtraction used in this paper.

The models for the extended MIR emission did not include radiation heating at the inner disk edge or adiabatic heating of gravitational fragments in the disk (Paper I). These may enhance the MIR emission, particularly at long wavelengths such as $\lambda$=11.3 \micron.
This will be investigated in the future using more sophisticated radiative transfer simulations.


\begin{acknowledgements}
M.T. is supported by the National Science and Technology Council (NSTC) of Taiwan (grant No. 109-2112-M-001-019, 110-2112-M-001-044-, 111-2112-M-001-059-, 112-2112-M-001-031-, 113-2112-M-001-009).
M.T. was also supported by NSTC Taiwan 108-2923-M-001-006-MY3 for the Taiwanese-Russian collaboration project.
This work has made use of data from the European Space Agency (ESA) mission Gaia (https://www.cosmos.esa.int/gaia), processed by the Gaia Data Processing and Analysis Consortium (DPAC, https://www.cosmos.esa.int/web/gaia/dpac/consortium). Funding for the DPAC has been provided by national institutions, in particular the institutions participating in the Gaia Multilateral Agreement.
This research made use of the Simbad database operated at CDS, Strasbourg, France, and the NASA's Astrophysics Data System Abstract Service.
\end{acknowledgements}

\vspace{5mm}
\facilities{ELT (METIS)}
\software{
HEEPS \citep{Delacroix22},
numpy \citep{numpy},
scipy \citep{scipy},
astropy \citep{astropy}, 
          }



\bibliographystyle{aasjournal}


\appendix

\section{Simulated Images without PSF subtraction} \label{app:no_psfsub}

\begin{figure*}[ht!]
\centering
\includegraphics[width=15cm]{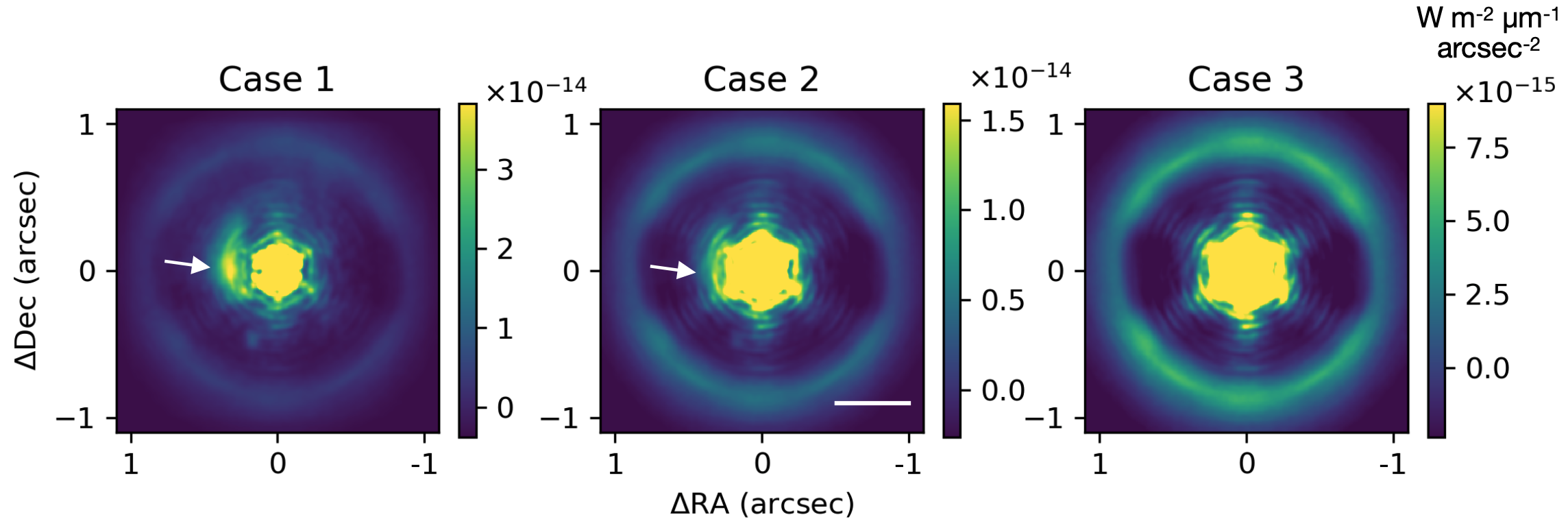}
\caption{
Same as Figure \ref{fig:FU_4} but for $\lambda$=3.8 \micron~only, Cases 1-3 and without subtracting the bright central emission.
The color contrast for each image is arbitrarily adjusted to investigate how the observations of the extended emission suffer from the emission from the central disk. The white arrows indicate the eastern arm in the extended emission shown in Figures \ref{fig:PI_H} and \ref{fig:FU_4}.
The horizontal bar at the bottom-right of the middle panel indicates a spatial scale of 200 au.
\label{fig:no_psfsub}
}
\end{figure*}

In this section, we limit our discussion to $\lambda$=3.8 \micron, which yields the best detection of the extended emission (Section 3.2, Figures \ref{fig:FU_4} and \ref{fig:V17_4}).
Figure \ref{fig:no_psfsub} shows the simulated images for FU Ori for three cases (Case 1-3; see Table 5) without subtracting the bright central emission.
The east arm shown in Figures \ref{fig:PI_H} and \ref{fig:FU_4} is visible for Case 1, i.e., for which we expect the observations with the best signal-to-noise (see Section 3.2, Figure \ref{fig:FU_4}, and Table \ref{tbl:cases}).
However, this arm is only marginally visible in Case 2 and not clearly visible in Case 3. These contrast with Figure \ref{fig:FU_4} for the same cases but after subtracting the bright central emission, for which the arm is clearly observed. As such, the subtraction of the bright central emission significantly enhances the detection of the extended emission.

\section{Exposure time and detector saturation} \label{app:saturation}

\begin{figure*}[ht!]
\centering
\includegraphics[width=18cm]{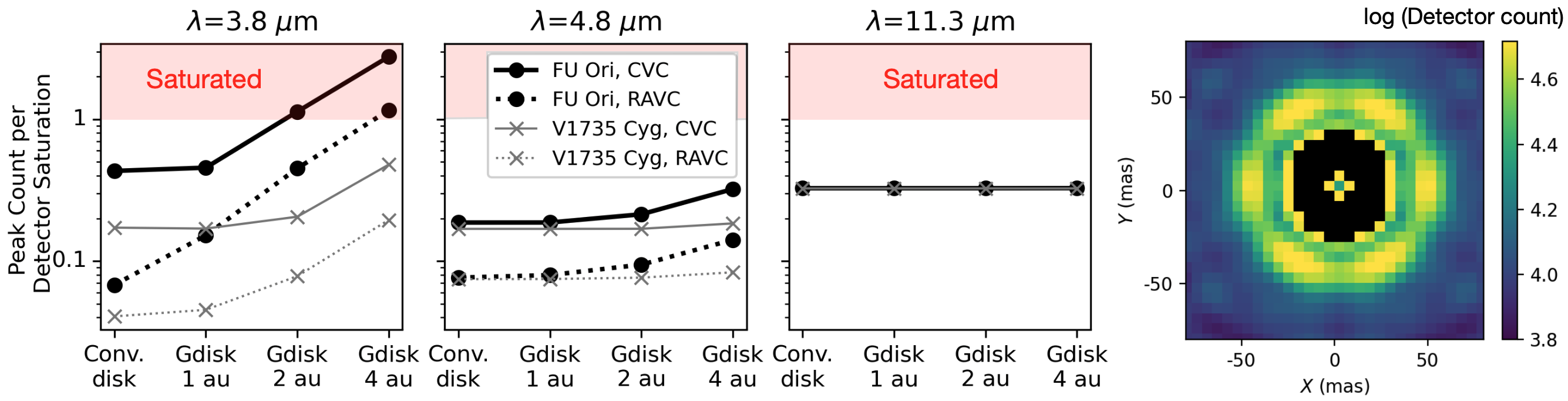}
\caption{
({\it left to middle}) The peak counts per the level of the detector saturation for various cases.
The panels from left to right are for three wavelengths of the observations.
In each panel, the horizontal axis shows the cases where the central source is a point source, a conventional accretion disk, and Gaussian disks with a HWHM of 1, 2, and 4 au, respectively.
The plots for the two target objects completely overlap for $\lambda$=11.3 \micron~in the right panel.
({\it right}) The image of the central source for the most saturated case, i.e., FU Ori, where the central source is a Gaussian disk with a HWHM of 4 au, observed using the CVC coronagraph at $\lambda$=3.8 \micron.
The black area at the center indicates the region where the detector saturation occurred.
\label{fig:saturation}
}
\end{figure*}

The left-to-middle panels in Figure \ref{fig:saturation} show the peak intensities divided by the saturation levels of the detector for the individual cases. For the observations at $\lambda$=3.8 \micron, we selected an exposure of 0.1 second, that is an approximate minimum exposure for which the thermal background photon noise dominates over the read noise of the detector.
For those at $\lambda$=4.8 and 11.3 \micron, we selected the minimum exposure times expected to be accepted for the detector, for which the read noise of the detectors provides only a fraction of the total noise.  See Table \ref{tbl:inst} for the details of these exposures and related parameters.

At $\lambda$=3.8 \micron, the peak count increases with the spatial distribution of the central disk. This trend is marginal and absent at $\lambda$=4.8 and 11.3 \micron,
respectively, due to the significantly larger thermal background. The figure shows that the detector saturation occurred for only a few limited cases for FU Ori observed using the CVC coronagraph. The right panel of Figure \ref{fig:saturation} shows the most saturated case (for a Gaussian disk with a HWHM of 4 au), for which the detector saturation occurs only at the angular scale of the diffraction core.

Throughout the paper, we assumed that the detector saturation does not significantly degrade the accuracy for aligning the image when we subtract the PSF using a reference star. We also note that the bright central emission of FU Ori at $\lambda$=3.8 \micron~is not likely due to a disk with the largest spatial extent (a Gaussian disk with a HWHM of 4 au), which caused significant detector saturation (Section \ref{sec:method:cs}).

\section{Complementary simulations with the RAVC coronagraph} \label{app:psf_ravc}

\begin{figure*}[ht!]
\centering
\includegraphics[width=17cm]{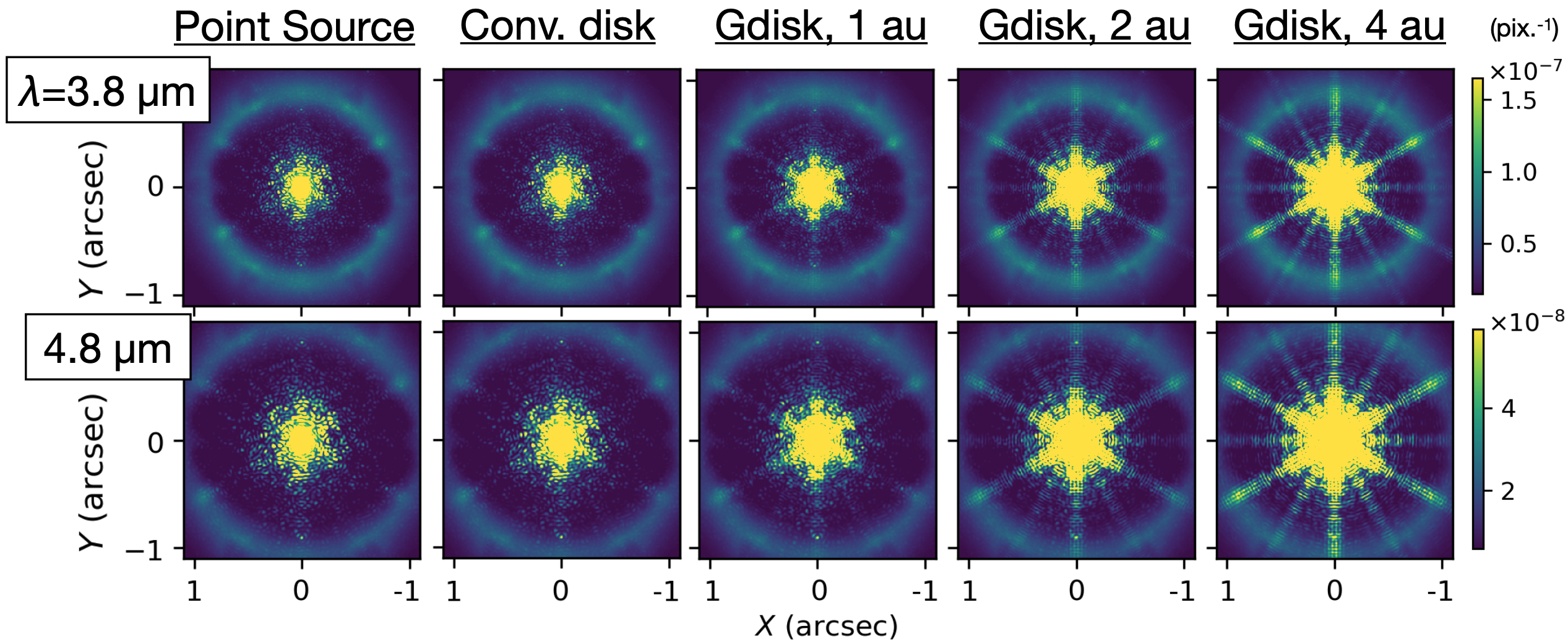}
\caption{
Same as Figure \ref{fig:psfs_cvc} but for the RAVC coronagraph.
\label{fig:psfs_ravc}
}
\end{figure*}

Figure \ref{fig:psfs_ravc} shows the PSFs for various cases simulated for the RAVC coronagraph. These qualitatively exhibit the same trends as the CVC coronagraph described in Section \ref{sec:results:basic} but at lower intensity levels.

\begin{figure}[ht!]
\centering
\includegraphics[width=9cm]{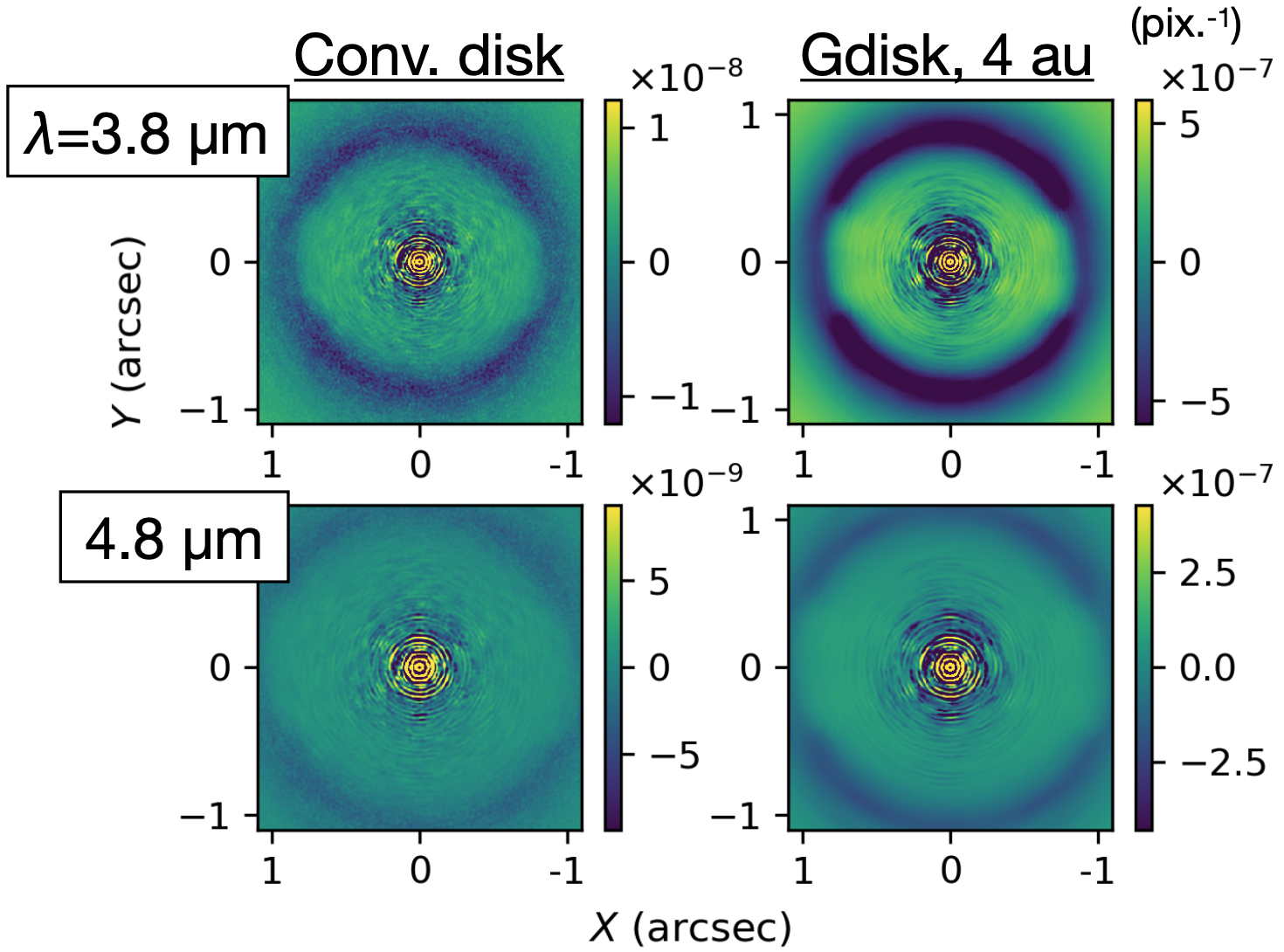}
\caption{
Same as Figure \ref{fig:psfsub_cvc} but for the RAVC coronagraph. The images are shown for the smallest and largest central disks (i.e., the conventional disk and a Gaussian disk with a HWHM of 4 au).
\label{fig:psfsub_ravc}
}
\end{figure}

Figure \ref{fig:psfsub_ravc} shows the on-axis PSFs for FU Ori after subtracting the reference PSF, i.e., the same figure as Figure \ref{fig:psfsub_cvc} but for the RAVC coronagraph. Due to the similarity of the images between different central disks (the conventional accretion disk and the Gaussian disks with a HWHM of 1, 2, and 4 au), we show those only for the smallest and largest disks, i.e., the conventional accretion disk and the Gaussian disks with a HWHM of 4 au, respectively.
As in Figure \ref{fig:psfsub_cvc}, the images for the different central disks are similar, but the absolute level of the speckle noise is associated with the spike patterns, and a negative ring occurs due to the flux scaling of the reference PSF.
Compared with Figure \ref{fig:psfs_ravc}, the spike patterns became significantly less noticeable because of (1) the PSF subtraction and (2) the field derotation of the individual frames before stacking.

\end{document}